\documentclass[preprint,prd,aps,showpacs,showkeys,nofootinbib]{revtex4}
\usepackage{graphicx}
\begin{document}

Published in JHEP 05 (2023) 075

\title{GKZ hypergeometric systems of the three-loop vacuum Feynman integrals}

\author{
Hai-Bin Zhang$^{1,2,3,}$\footnote{hbzhang@hbu.edu.cn},
Tai-Fu Feng$^{1,2,3,4,5,}$\footnote{fengtf@hbu.edu.cn}
}

\affiliation{
$^1$Department of Physics, Hebei University, Baoding, 071002, China\\
$^2$Hebei Key Laboratory of High-precision Computation and Application of
Quantum Field Theory, Baoding, 071002, China\\
$^3$Research Center for Computational Physics of Hebei Province, Baoding, 071002, China\\
$^4$Department of Physics, Guangxi University, Nanning, 530004, China\\
$^5$Department of Physics, Chongqing University, Chongqing, 401331, China
}

\begin{abstract}
We present the Gel'fand-Kapranov-Zelevinsky (GKZ) hypergeometric systems of the Feynman integrals of the three-loop vacuum diagrams with arbitrary masses, basing on Mellin-Barnes representations and Miller's transformation. The codimension of derived GKZ hypergeometric systems equals the number of independent dimensionless ratios among the virtual masses squared. Through GKZ hypergeometric systems, the analytical hypergeometric series solutions can be obtained in neighborhoods of origin including infinity. The linear independent hypergeometric series solutions whose convergent regions have non-empty intersection can constitute a fundamental solution system in a proper subset of the whole parameter space. The analytical expression of the vacuum  integral can be formulated as a linear combination of the corresponding fundamental solution system in certain convergent region.
\end{abstract}


\maketitle

\newpage
\tableofcontents

\newpage

\section{Introduction\label{sec1}}
\indent\indent

The higher-order radiative corrections are more important, with the increasing precision of measurements at the planned future colliders \cite{CLIC,ILC,CEPC,FCC,HL-LHC,Heinrich2021}. Compared to the experimental and parametric errors, it is useful to reduce corresponding theoretical uncertainties for precision test of the Standard Model (SM). With the improvement of experimental measurement accuracy, Feynman integrals need to be calculated beyond two-loop order. In this article, we investigate the analytical calculation for the Feynman integrals of the three-loop vacuum diagrams with arbitrary masses.

The completely general one-loop integrals are well known analytically in the time-space dimension $D=4-2\varepsilon$ \cite{tHooft1979,Passarino1979,Denner1993,V.A.Smirnov2012}.
At the two-loop level, the general vacuum integrals have been calculated to polylogarithms or
equivalent functions \cite{R2loop1,R2loop2,R2loop3,R2loop4,R2loop5}. But, the vacuum integrals at the three-loop level are only calculated analytically for one and two independent mass scales
\cite{R3loop1,R3loop2,R3loop3,R3loop4,R3loop5,R3loop6,R3loop7,R3loop8,R3loop9,R3loop10,R3loop11,R3loop12,R3loop13,R3loop14,R3loop15,R3loop16}.
Of course, some numerical solutions of general three-loop vacuum integrals have been obtained \cite{R3loopN1,R3loopN2,R3loopN3}.
Recently, using auxiliary mass flow numerical method \cite{AMFlow1,AMFlow2,AMFlow3,AMFlow4}, Feynman integrals can be reduced to the vacuum integrals which can be numerical solved by further reduction.
In order to further improve the computational efficiency and give analytical results completely, it is meaningful to explore new analytical calculating method of the three-loop vacuum integrals with arbitrary masses.

During the past decades, Feynman integrals have long been considered as the generalized hypergeometric functions~\cite{Regge1967,Davydychev1,Davydychev3,Davydychev1991JMP,Davydychev1992JPA,Davydychev1992JMP,
Davydychev1993,Berends1994,Smirnov1999,Tausk1999,Davydychev2000,Tarasov2000,Tarasov2003,Davydychev2006,Kalmykov2009,
Kalmykov2011,Kalmykov2012,Bytev2015,Bytev2016,Kalmykov2017,Feng2018,Feng2019,Gu2019,Gu2020,
Ananthanarayan2020,Ananthanarayan2021}.
Considering Feynman integrals as the generalized hypergeometric functions,
one finds that the $D-$module of a Feynman diagram~\cite{Kalmykov2012,Nasrollahpoursamami2016} is isomorphic to
Gel'fand-Kapranov-Zelevinsky (GKZ) $D-$module~\cite{Gelfand1987,Gelfand1988,Gelfand1988a,Gelfand1989,Gelfand1990}.
GKZ-hypergeometric systems of Feynman integrals with codimension$=0,\;1$
are presented in Refs.~\cite{Cruz2019,Klausen2019}
through Lee-Pomeransky parametric representations~\cite{Lee2013}.
To construct canonical series solutions with suitable independent variables,
one should compute the restricted $D$-module of GKZ-hypergeometric system originating from
Lee-Pomeransky representations on corresponding hyperplane in the parameter
space~\cite{Oaku1997,Walther1999,Oaku2001}.
In our previous work, GKZ hypergeometric systems of one- and two-loop Feynman diagrams
are obtained~\cite{Feng2020,GKZ-2loop,Grassmannians} from Mellin-Barnes representations~\cite{Feng2018,Feng2019}, through Miller's transformation~\cite{Miller68,Miller72}. There are some recent work in GKZ framework of Feynman integrals
\cite{Loebbert2020,Klemm2020,Bonisch2021,Hidding2021,Borinsky2020,Kalmykov2021,Tellander2021,Klausen2021,Mizera2021,
Arkani-Hamed2022,Chestnov2022,Walther2022,Ananthanarayan2022GKZ}.

In this article, we derive GKZ hypergeometric systems of the Feynman integrals of the three-loop vacuum diagrams with arbitrary masses, basing on Mellin-Barnes representations and Miller's transformation. The generally strategy for analyzing the three-loop vacuum integrals includes three steps here.
Firstly, we obtain the Mellin-Barnes representation of the vacuum integral. Secondly, we find GKZ hypergeometric system of the vacuum integral via Miller's transformation.
Finally,  analytical  hypergeometric series solutions of the vacuum integral are constructed  in neighborhoods of origin including infinity.
The integration constants, i.e. the combination coefficients, are determined from the vacuum integral of an ordinary point or some regular singularities.

Our presentation is organized as following. Through the Mellin-Barnes representation  and Miller's transformation,
we derive the GKZ hypergeometric system of Feynman integral of the
three-loop vacuum diagram with four propagates in Sec. \ref{sec2}.
And then, we construct the  analytical  hypergeometric series solutions of the GKZ system of the three-loop vacuum integral with four propagates in Sec. \ref{sec3}.
We also derive the GKZ hypergeometric system of Feynman integral of the
three-loop vacuum diagram with five propagates in Sec. \ref{sec-five}.
We elucidate how to obtain the analytical expression clearly in Sec. \ref{sec-spe}, assuming the two nonzero virtual mass for the three-loop vacuum integral with five propagates.
At last, the conclusions are summarized in Sec. \ref{sec-con}, and some formulates
are presented in the appendices.

\section{GKZ hypergeometric system of the three-loop vacuum integral with four propagates\label{sec2}}
\indent\indent

\begin{figure}[ht]
\setlength{\unitlength}{0cm}
\centering
\hspace{-1.5cm}\hspace{2cm}
\includegraphics[width=5.2cm]{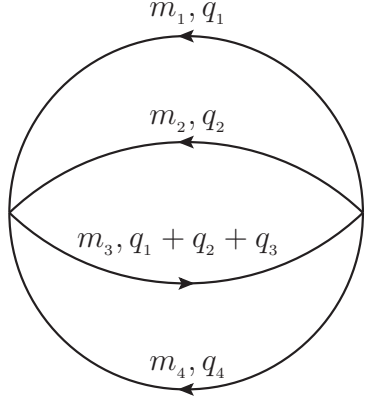}
\vspace{0cm}
\caption[]{Three-loop vacuum diagram with four propagators, which $m_{_i}$ denotes the mass of the $i$-th particle and $q_{_j}$ denotes the momentum.}
\label{fig-3loop4}
\end{figure}

The general analytic expression for the Feynman integral of the three-loop vacuum diagram with four propagates in Fig.~\ref{fig-3loop4} is written as
\begin{eqnarray}
&&U_{_4}=\Big(\Lambda_{_{\rm RE}}^2\Big)^{6-\frac{3D}{2}}\int{d^Dq_{_1}\over(2\pi)^D}
{d^Dq_{_2}\over(2\pi)^D}{d^Dq_{_3}\over(2\pi)^D}
\nonumber\\
&&\hspace{1.1cm}\times
{1\over(q_{_1}^2-m_{_1}^2)(q_{_2}^2-m_{_2}^2)((q_{_1}+q_{_2}+q_{_3})^2-m_{_3}^2)
(q_{_3}^2-m_{_4}^2)}\:,
\label{3loop-1}
\end{eqnarray}
where $D=4-2\varepsilon$ is the number of dimensions in dimensional regularization and $\Lambda_{_{\rm RE}}$ denotes the renormalization energy scale.
Adopting the notation of Refs.~\cite{Feng2018,Feng2019}, the Feynman integral of the three-loop vacuum diagram with four propagates can be written as
\begin{eqnarray}
&&U_{_4}=\frac{\Big(\Lambda_{_{\rm RE}}^2\Big)^{6-\frac{3D}{2}}}{(2\pi i)^3} \int_{-i\infty}^{+i\infty}ds_{_1}ds_{_2}ds_{_3}
\Big[\prod\limits_{i=1}^3(-m_{_i}^2)^{s_{_i}}\Gamma(-s_{_i})\Gamma(1+s_{_i})\Big]I_{q}\:,
\label{3loop-2}
\end{eqnarray}
where
\begin{eqnarray}
I_{q}&&\equiv \int{d^Dq_{_1}\over(2\pi)^D}{d^Dq_{_2}\over(2\pi)^D} {d^Dq_{_3}\over(2\pi)^D}
{1\over(q_{_1}^2)^{1+s_{_1}}(q_{_2}^2)^{1+s_{_2}}((q_{_1}+q_{_2}+q_{_3})^2)^{1+s_{_3}}
(q_{_3}^2-m_{_4}^2)}\:.
\label{3loop-2-1}
\end{eqnarray}

First, we can integrate out $q_{_2}$:
\begin{eqnarray}
I_{q}&&={\Gamma(2+s_{_2}+s_{_3})\over\Gamma(1+s_{_2})\Gamma(1+s_{_3})}
\int{d^Dq_{_1}\over(2\pi)^D}{d^Dq_{_3}\over(2\pi)^D}
{1\over(q_{_1}^2)^{1+s_{_1}}(q_{_3}^2-m_{_4}^2)}
\nonumber\\
&&\hspace{0.4cm}\times
\int_0^1dx(1-x)^{s_{_2}}x^{s_{_3}}\int{d^Dq_{_2}\over(2\pi)^D}
{1\over[q_{_2}^2+x(1-x)(q_{_1}+q_{_3})^2]^{2+s_{_2}+s_{_3}}}\:.
\label{3loop-3}
\end{eqnarray}
Using the well-known integral
\begin{eqnarray}
\int{d^Dq\over(2\pi)^D}{1\over[q^2+{\triangle}]^n}
={i(-)^{D/2}\Gamma(n-{D\over2})\over (4\pi)^{D/2}\Gamma(n)}{\Big( {1\over {\triangle}} \Big)}^{n-D/2}\:,
\label{integral}
\end{eqnarray}
one can have
\begin{eqnarray}
I_{q}&&=
{i(-)^{D/2}\Gamma(2-{D\over2}+s_{_2}+s_{_3})\over(4\pi)^{D/2}\Gamma(1+s_{_2})\Gamma(1+s_{_3})}
\int{d^Dq_{_1}\over(2\pi)^D}{d^Dq_{_3}\over(2\pi)^D}
{1\over(q_{_1}^2)^{1+s_{_1}}((q_{_1}+q_{_3})^2)^{2-{D\over2}+s_{_2}+s_{_3}}(q_{_3}^2-m_{_4}^2)}
\nonumber\\
&&\hspace{0.4cm}\times
\int_0^1dx(1-x)^{D/2-2-s_{_3}}x^{D/2-2-s_{_2}}\:.
\label{3loop-3-1}
\end{eqnarray}
Through Beta function
\begin{eqnarray}
B(m,n)=\int_0^1dx\: x^{m-1}(1-x)^{n-1} ={ \Gamma(m)\Gamma(n)\over \Gamma(m+n)}\:,
\label{Beta}
\end{eqnarray}
we can have
\begin{eqnarray}
I_{q}&&=
{i(-)^{D/2}\Gamma(2-{D\over2}+s_{_2}+s_{_3})\Gamma({D\over2}-1-s_{_2})\Gamma({D\over2}-1-s_{_3})
\over(4\pi)^{D/2}\Gamma(1+s_{_2})\Gamma(1+s_{_3})\Gamma(D-2-s_{_2}-s_{_3})}
\nonumber\\
&&\hspace{0.4cm}\times
\int{d^Dq_{_1}\over(2\pi)^D}{d^Dq_{_3}\over(2\pi)^D}
{1\over(q_{_1}^2)^{1+s_{_1}}((q_{_1}+q_{_3})^2)^{2-{D\over2}+s_{_2}+s_{_3}}(q_{_3}^2-m_{_4}^2)}\:.
\label{3loop-4}
\end{eqnarray}

Second, we can integrate out $q_{_1}$:
\begin{eqnarray}
&&\int{d^Dq_{_1}\over(2\pi)^D}{d^Dq_{_3}\over(2\pi)^D}
{1\over(q_{_1}^2)^{1+s_{_1}}((q_{_1}+q_{_3})^2)^{2-{D\over2}+s_{_2}+s_{_3}}(q_{_3}^2-m_{_4}^2)}
\nonumber\\
&&\hspace{-0.0cm}=
{i(-)^{D/2}\Gamma(3-D+\sum\limits_{i=1}^3s_{_i})\Gamma({D\over2}-1-s_{_1})\Gamma(D-2-s_{_2}-s_{_3})
\over(4\pi)^{D/2}\Gamma(1+s_{_1})\Gamma(2-{D\over2}+s_{_2}+s_{_3})
\Gamma({3D\over2}-3-\sum\limits_{i=1}^3s_{_i})}
\nonumber\\
&&\hspace{0.4cm}\times
\int{d^Dq_{_3}\over(2\pi)^D}
{1\over(q_{_3}^2-m_{_4}^2)(q_{_3}^2)^{3-D+s_{_1}+s_{_2}+s_{_3}}}\:.
\label{3loop-5}
\end{eqnarray}

Third, we integrate out $q_{_3}$:
\begin{eqnarray}
&&\int{d^Dq_{_3}\over(2\pi)^D}
{1\over(q_{_3}^2-m_{_4}^2)(q_{_3}^2)^{3-D+s_{_1}+s_{_2}+s_{_3}}}
\nonumber\\
&&\hspace{-0.0cm}={{i} \over (4\pi)^{D/2}} (-)^{4-D+\sum\limits_{i=1}^3s_{_i}} \Big( {1\over m_{_4}^2}\Big)^{4-{3D\over2}+\sum\limits_{i=1}^3s_{_i}}
{\Gamma(4-{3D\over2}+\sum\limits_{i=1}^3s_{_i}) \Gamma({3D\over2}-3-\sum\limits_{i=1}^3s_{_i}) }\:.
\label{3loop-6}
\end{eqnarray}

Together with Eqs. (\ref{3loop-4}-\ref{3loop-6}), one can have
\begin{eqnarray}
I_{q}&&={{-i} \over (4\pi)^{3D\over2}} (-)^{\sum\limits_{i=1}^3s_{_i}} \Big( {1\over m_{_4}^2}\Big)^{4-{3D\over2}+\sum\limits_{i=1}^3s_{_i}}
\Big[\prod\limits_{i=1}^3\Gamma({D\over2}-1-s_{_i})\Gamma(1+s_{_i})^{-1}\Big]
\nonumber\\
&&\hspace{0.5cm}\times
\Gamma(3-D+\sum\limits_{i=1}^3s_{_i})
\Gamma(4-{3D\over2}+\sum\limits_{i=1}^3s_{_i})\;.
\label{3loop-7-1}
\end{eqnarray}
And then, the Mellin-Barnes representation of the Feynman integral of the three-loop vacuum diagram in Eq. (\ref{3loop-2}) can be written as
\begin{eqnarray}
&&U_{_4}=
{-im_{_4}^4\over(2\pi i)^3(4\pi)^6}\Big({4\pi\Lambda_{_{\rm RE}}^2\over m_{_4}^2}\Big)^{6-{3D\over2}}
\int_{-i\infty}^{+i\infty}ds_{_1}ds_{_2}ds_{_3}
\Big[\prod\limits_{i=1}^3\Big({m_{_i}^2\over m_{_4}^2}\Big)^{s_{_i}}\Gamma(-s_{_i})\Big]
\nonumber\\
&&\hspace{1.0cm}\times
\Big[\prod\limits_{i=1}^3\Gamma({D\over2}-1-s_{_i})\Big]\Gamma(3-D+\sum\limits_{i=1}^3s_{_i})
\Gamma(4-{3D\over2}+\sum\limits_{i=1}^3s_{_i})\;.
\label{3loop-7}
\end{eqnarray}

It is well known that negative integers and zero are simple poles of the function
$\Gamma(z)$. As all $s_{_i}$ contours are closed to the right in corresponding
complex planes, one finds that the analytic expression of the three-loop vacuum  integral
can be written as the linear combination of generalized hypergeometric functions. Taking the residue of the pole of $\Gamma(-s_{_i}),\;(i=1,\;2,\;3)$, we can derive one linear independent term:
\begin{eqnarray}
&&U_{_4}\ni
{-im_{_4}^4\over(4\pi)^6}\Big({4\pi\Lambda_{_{\rm RE}}^2\over m_{_4}^2}\Big)^{6-{3D\over2}}
\sum\limits_{n_{_1}=0}^\infty\sum_{n_{_2}=0}^\infty\sum_{n_{_3}=0}^\infty
(-)^{\sum\limits_{i=1}^3n_{_i}}x_{_1}^{n_{_1}}x_{_2}^{n_{_2}}x_{_3}^{n_{_3}}
\nonumber\\
&&\hspace{1.0cm}\times
\Big[\prod\limits_{i=1}^3\Gamma({D\over2}-1-n_{_i})(n_{_i}!)^{-1}\Big]\Gamma(3-D+\sum\limits_{i=1}^3n_{_i})
\Gamma(4-{3D\over2}+\sum\limits_{i=1}^3n_{_i})\:,
\label{3loop-8}
\end{eqnarray}
with $x_{_i}={m_{_i}^2\over m_{_4}^2},\;(i=1,\;2,\;3)$.

We adopt the identity
\begin{eqnarray}
&&\Gamma(z-n)\Gamma(1-z+n)=(-)^{n}\Gamma(z)\Gamma(1-z)=(-)^{n}\pi/\sin\pi z\:,
\label{Gamma}
\end{eqnarray}
which originate from the well-known relation $\Gamma(z)\Gamma(1-z)=\pi/\sin\pi z$.
Then, Eq.~(\ref{3loop-8}) can be written as
\begin{eqnarray}
U_{_4}\ni
{im_{_4}^4\over(4\pi)^6}\Big({4\pi\Lambda_{_{\rm RE}}^2\over m_{_4}^2}\Big)^{6-{3D\over2}}
{ \pi^3\over \sin^3\frac{\pi D}{2} }
T_{_{4}}({\bf a},\;{\bf b}\;\Big|\;{\bf x})\;,
\label{3loop-9}
\end{eqnarray}
with
\begin{eqnarray}
T_{_{4}}({\bf a},\;{\bf b}\;\Big|\;{\bf x})=\sum\limits_{n_{_1}=0}^\infty\sum_{n_{_2}=0}^\infty\sum_{n_{_3}=0}^\infty
A_{_{n_{_1}n_{_2}n_{_3}}} x_{_1}^{n_{_1}}x_{_2}^{n_{_2}}x_{_3}^{n_{_3}}\:,
\label{3loop-10}
\end{eqnarray}
where ${\bf x}=(x_{_1},x_{_2},x_{_3})$, ${\bf a}=(a_{_1},a_{_2})$ and ${\bf b}=(b_{_1},b_{_2},b_{_3})$ with
\begin{eqnarray}
&&a_{_1}=3-D,\;a_{_2}=4-{3D\over2},\;b_{_1}=b_{_2}=b_{_3}=2-{D\over2}\:,
\label{3loop-11}
\end{eqnarray}
and the coefficient $A_{_{n_{_1}n_{_2}n_{_3}}}$ is
\begin{eqnarray}
\hspace{-0.5cm}A_{_{n_{_1}n_{_2}n_{_3}}}=
{\Gamma(a_{_1}+\sum\limits_{i=1}^3n_{_i}) \Gamma(a_{_2}+\sum\limits_{i=1}^3n_{_i})
\over  n_{_1}!n_{_2}!n_{_3}!\Gamma(b_{_1}+n_{_1})\Gamma(b_{_2}+n_{_2})\Gamma(b_{_3}+n_{_3})}\:.
\label{3loop-12}
\end{eqnarray}

Through the adjacent relations of the coefficient $A_{_{n_{_1}n_{_2}n_{_3}}}$:
\begin{eqnarray}
&&{A_{_{(n_{_1}+1)n_{_2}n_{_3}}}\over A_{_{n_{_1}n_{_2}n_{_3}}}}=
{(a_{_1}+\sum\limits_{i=1}^3n_{_i})(a_{_2}+\sum\limits_{i=1}^3n_{_i})
\over(n_{_1}+1)(b_{_1}+n_{_1})}
\;,\nonumber\\
&&{A_{_{n_{_1}(n_{_2}+1)n_{_3}}}\over A_{_{n_{_1}n_{_2}n_{_3}3}}}=
{(a_{_1}+\sum\limits_{i=1}^3n_{_i})(a_{_2}+\sum\limits_{i=1}^3n_{_i})
\over (n_{_2}+1)(b_{_2}+n_{_2})}
\;,\nonumber\\
&&{A_{_{n_{_1}n_{_2}(n_{_3}+1)}}\over A_{_{n_{_1}n_{_2}n_{_3}}}}=
{(a_{_1}+\sum\limits_{i=1}^3n_{_i})(a_{_2}+\sum\limits_{i=1}^3n_{_i})
\over (n_{_3}+1)(b_{_3}+n_{_3})}\;,
\label{3loop-13}
\end{eqnarray}
the difference-differential operators are written as
\begin{eqnarray}
&&(\sum\limits_{i=1}^3\vartheta_{_{x_{_i}}}+a_{_j})T_{_{4}}({\bf a},\;{\bf b}\;\Big|\;{\bf x})
=a_{_j}T_{_{4}}({\bf a}+{\bf e}_{_{2,j}},\;{\bf b}\;\Big|\;{\bf x})\;,\;(j=1,2)
\;,\nonumber\\
&&(\vartheta_{_{x_{_k}}}+b_{_k}-1)T_{_{4}}({\bf a},\;{\bf b}\;\Big|\;{\bf x})
=(b_{_k}-1)T_{_{4}}({\bf a},\;{\bf b}-{\bf e}_{_{3,k}}\;\Big|\;{\bf x})\;,
\nonumber\\
&&\partial_{_{x_{_k}}}T_{_{4}}({\bf a},\;{\bf b}\;\Big|\;{\bf x})={a_{_1}a_{_2}\over b_{_k}}
T_{_{4}}({\bf a}+{\bf e}_{_{2}},\;{\bf b}+{\bf e}_{_{3,k}}\;\Big|\;{\bf x})\;
,\;(k=1,2,3)\;.
\label{3loop-14}
\end{eqnarray}
Here ${\bf e}_{_{2,j}}\in{\bf R}^2\;(j=1,2)$ with ${\bf e}_{_{2,1}}=(1,\;0)$ and ${\bf e}_{_{2,2}}=(0,\;1)$, ${\bf e}_{_{2}}=(1,\;1)$, ${\bf e}_{_{3,k}}\in{\bf R}^3\;(k=1,2,3)$ denotes the row vector whose entry is zero except that the $k-$th entry is $1$, ${\bf e}_{_{3}}=(1,\;1,\;1)$, $\vartheta_{_{x_{_k}}}=x_{_k}\partial_{_{x_{_k}}}$
denotes the Euler operators, and $\partial_{_{x_{_k}}}=\partial/\partial x_{_k}$, respectively.

We can define the auxiliary function
\begin{eqnarray}
&&\Phi_{_{4}}({\bf a},\;{\bf b}\;\Big|\;{\bf x},\;{\bf u},\;{\bf v})={\bf u}^{\bf a}{\bf v}^{{\bf b}-{\bf e}_{_{3}}}T_{_{4}}({\bf a},\;{\bf b}\;\Big|\;{\bf x})\;,
\label{3loop-15}
\end{eqnarray}
with the intermediate variables ${\bf u}=(u_1,\;u_2)=(1,\;1)$, ${\bf v}=(v_1,\;v_2,\;v_3)=(1,\;1,\;1)$.
Through Miller's transformation \cite{Miller68,Miller72}, the relations are obtained
\begin{eqnarray}
&&\vartheta_{_{u_j}}\Phi_{_{4}}({\bf a},\;{\bf b}\;\Big|\;{\bf x},\;{\bf u},\;{\bf v})=a_{_j}
\Phi_{_{4}}({\bf a},\;{\bf b}\;\Big|\;{\bf x},\;{\bf u},\;{\bf v})
\;,\;(j=1,2)\:,\nonumber\\
&&\vartheta_{_{v_k}}\Phi_{_{4}}({\bf a},\;{\bf b}\;\Big|\;{\bf x},\;{\bf u},\;{\bf v})=(b_{_k}-1)
\Phi_{_{4}}({\bf a},\;{\bf b}\;\Big|\;{\bf x},\;{\bf u},\;{\bf v})\;,\;(k=1,2,3)\:,
\label{3loop-16}
\end{eqnarray}
which naturally induces the notion of GKZ hypergeometric system.

In addition, the contiguous relations of Eq.~(\ref{3loop-14}) are rewritten as
\begin{eqnarray}
&&u_{_j}(\sum\limits_{i=1}^4\vartheta_{_{x_{_i}}}+\vartheta_{_{u_j}}) \Phi_{_{4}}({\bf a},\;{\bf b}\;\Big|\;{\bf x},\;{\bf u},\;{\bf v})
=a_{_j}\Phi_{_{4}}({\bf a}+{\bf e}_{_{2,j}},\;{\bf b}\;\Big|\;{\bf x},\;{\bf u},\;{\bf v}),\;
\nonumber\\
&&{1\over v_{_k}}(\vartheta_{_{x_{_k}}}+\vartheta_{_{v_k}})\Phi_{_{4}}({\bf a},\;{\bf b}\;\Big|\;{\bf x},\;{\bf u},\;{\bf v})
=(b_{_k}-1)\Phi_{_{4}}({\bf a},\;{\bf b}-{\bf e}_{_{3,k}}\;\Big|\;{\bf x},\;{\bf u},\;{\bf v}),\;
\nonumber\\
&&u_{_1}u_{_2}v_{_k}\partial_{_{x_{_k}}}\Phi_{_{4}}({\bf a},\;{\bf b}\;\Big|\;{\bf x},\;{\bf u},\;{\bf v})={a_{_1}a_{_2}\over b_{_k}}
\Phi_{_{4}}({\bf a}+{\bf e}_{_{2}},\;{\bf b}+{\bf e}_{_{3,k}}\;\Big|\;{\bf x},\;{\bf u},\;{\bf v})
\;.
\label{3loop-17}
\end{eqnarray}
The operators above together with $\vartheta_{_{u_j}}$, $\vartheta_{_{v_j}}$
define the Lie algebra of the hypergeometric systems~\cite{Miller68,Miller72}.
Through the transformation
\begin{eqnarray}
&&z_{_j}={1\over u_{_j}},\;(j=1,2),\;\;z_{_{2+k}}=v_{_k},\;z_{_{5+k}}={x_{_k}\over u_{_1}u_{_2}v_{_k}},\;(k=1,2,3),
\label{3loop-18}
\end{eqnarray}
one derives
\begin{eqnarray}
&&\vartheta_{_{u_{_1}}}=-\vartheta_{_{z_{_1}}}-\vartheta_{_{z_{_{6}}}}-\vartheta_{_{z_{_{7}}}}
-\vartheta_{_{z_{_{8}}}}\;,\nonumber\\
&&\vartheta_{_{u_{_2}}}=-\vartheta_{_{z_{_2}}}-\vartheta_{_{z_{_{6}}}}-\vartheta_{_{z_{_{7}}}}
-\vartheta_{_{z_{_{8}}}},
\nonumber\\
&&\vartheta_{_{v_{_k}}}=\vartheta_{_{z_{_{2+k}}}}-\vartheta_{_{z_{_{5+k}}}},\;\;
\vartheta_{_{x_{_k}}}=\vartheta_{_{z_{_{5+k}}}}\:.
\label{3loop-19}
\end{eqnarray}

Through Eq.~(\ref{3loop-16}), one can have the GKZ hypergeometric system for the three-loop vacuum integral  with four propagates
\begin{eqnarray}
&&\mathbf{A_{_{4}}}\cdot\vec{\vartheta}_{_{4}}\Phi_{_{4}}=\mathbf{B_{_{4}}}\Phi_{_{4}}\;,
\label{3loop-20}
\end{eqnarray}
where
\begin{eqnarray}
&&\mathbf{A_{_{4}}}=\left(\begin{array}{cccccccc}
1\;\;&0\;\;&0\;\;&0\;\;&0\;\;&1\;\;&1\;\;&1\;\;\\
0\;\;&1\;\;&0\;\;&0\;\;&0\;\;&1\;\;&1\;\;&1\;\;\\
0\;\;&0\;\;&1\;\;&0\;\;&0\;\;&-1\;\;&0\;\;&0\;\;\\
0\;\;&0\;\;&0\;\;&1\;\;&0\;\;&0\;\;&-1\;\;&0\;\;\\
0\;\;&0\;\;&0\;\;&0\;\;&1\;\;&0\;\;&0\;\;&-1\;\;\\
\end{array}\right)
\;,\nonumber\\
&&\vec{\vartheta}_{_{4}}^{\;T}=(\vartheta_{_{z_{_1}}},\;\cdots
,\;\vartheta_{_{z_{_{8}}}})
\;,\nonumber\\
&&\mathbf{B_{_{4}}}^{T}=(-a_{_1},\;-a_{_2},\;b_{_1}-1,\;b_{_2}-1
,\;b_{_3}-1)\;.
\label{3loop-21}
\end{eqnarray}
Here, the GKZ hypergeometric system for the three-loop vacuum integral with four propagates is in keeping with that for the two-loop sunset diagram \cite{Feng2020}.

Correspondingly the dual matrix $\mathbf{\tilde A_{_{4}}}$ of $\mathbf{A_{_{4}}}$ is
\begin{eqnarray}
&&\mathbf{\tilde A_{_{4}}}=\left(\begin{array}{cccccccc}
-1\;\;&-1\;\;&1\;\;&0\;\;&0\;\;&1\;\;&0\;\;&0\;\;\\
-1\;\;&-1\;\;&0\;\;&1\;\;&0\;\;&0\;\;&1\;\;&0\;\;\\
-1\;\;&-1\;\;&0\;\;&0\;\;&1\;\;&0\;\;&0\;\;&1\;\;
\end{array}\right).
\label{3loop-22}
\end{eqnarray}
The row vectors of the matrix $\mathbf{\tilde A_{_{4}}}$ induce the integer sublattice $\mathbf{B}$
which can be used to construct the formal solutions in hypergeometric series. Actually the integer sublattice $\mathbf{B}$ indicates that the solutions of the system should satisfy the equations in Eq. (\ref{3loop-20}) and the following hyperbolic equations simultaneously
\begin{eqnarray}
&&{\partial^2\Phi_{_{4}}\over\partial z_{_1}\partial z_{_2}}={\partial^2\Phi_{_{4}}\over\partial z_{_3}\partial z_{_{6}}}
\;,\nonumber\\
&&{\partial^2\Phi_{_{4}}\over\partial z_{_1}\partial z_{_2}}={\partial^2\Phi_{_{4}}\over\partial z_{_4}\partial z_{_{7}}}
\;,\nonumber\\
&&{\partial^2\Phi_{_{4}}\over\partial z_{_1}\partial z_{_2}}={\partial^2\Phi_{_{4}}\over\partial z_{_5}\partial z_{_{8}}}\;.
\label{3loop-23}
\end{eqnarray}

Actually those partial differential equations is a Gr\"obner basis of the toric ideal of the matrix
$\mathbf{\tilde A_{_{4}}}$ presented Eq.~(\ref{3loop-22}). Defining the combined variables
\begin{eqnarray}
y_{_1}={z_{_3}z_{_{6}}\over z_{_1}z_{_2}}
\;,\quad  y_{_2}={z_{_4}z_{_{7}}\over z_{_1}z_{_2}}
\;,\quad  y_{_3}={z_{_5}z_{_{8}}\over z_{_1}z_{_2}}
\;,
\label{3loop-24}
\end{eqnarray}
we write the solutions satisfying Eq. (\ref{3loop-20}) and Eq. (\ref{3loop-23}) as
\begin{eqnarray}
&&\Phi_{_{4}}({\mathbf z})=\Big(\prod\limits_{i=1}^{8}z_{_i}^{\alpha_{_i}}\Big)
\varphi_{_{4}}(y_{_1},\;y_{_2},\;y_{_3})\;.
\label{3loop-25}
\end{eqnarray}
Here $\vec{\alpha}^{\:T}=(\alpha_{_1},\;\alpha_{_2},\;\cdots,\;\alpha_{_{8}})$
denotes a sequence of complex number such that
\begin{eqnarray}
&&\mathbf{A_{_{4}}}\cdot\vec{\alpha}=\mathbf{B_{_{4}}}\;,
\label{3loop-26}
\end{eqnarray}
namely,
\begin{eqnarray}
&&\alpha_{_1}+\alpha_{_6}+\alpha_{_7}+\alpha_{_8}=-a_{_1}\;,
\quad  \alpha_{_2}+\alpha_{_6}+\alpha_{_7}+\alpha_{_8}=-a_{_2}\;,
\nonumber\\
&&\alpha_{_3}-\alpha_{_6}=b_{_1}-1\;,\quad \alpha_{_4}-\alpha_{_7}=b_{_2}-1\;,\quad \alpha_{_5}-\alpha_{_8}=b_{_3}-1\;.
\label{3loop-26-1}
\end{eqnarray}

Substituting Eq.~(\ref{3loop-25}) in Eq.~(\ref{3loop-23}), we obtain the following three
independent partial differential equations (PDEs):
\begin{eqnarray}
\hat{L}_{_i}\varphi_{_{4}}=0,\;\;(i=1,2,3),
\label{3loop-27}
\end{eqnarray}
where the linear partial differential operators $\hat{L}_{_i}$:
\begin{eqnarray}
&&\hat{L}_{_1}=(1-y_{_1})y_{_1}^2{\partial^2\over\partial y_{_1}^2}-y_{_1}\Big(y_{_2}^2{\partial^2\over\partial y_{_2}^2}+y_{_3}^2{\partial^2\over\partial y_{_3}^2}\Big)
-y_{_1}\Big(y_{_1}y_{_2}{\partial^2\over\partial y_{_1}\partial y_{_2}}+y_{_1}y_{_3}{\partial^2\over\partial y_{_1}\partial y_{_3}}
\nonumber\\
&&\hspace{1.1cm}
+y_{_2}y_{_3}{\partial^2\over\partial y_{_2}\partial y_{_3}}\Big)
+\Big[(1+\alpha_{_3}+\alpha_{_{6}})-(1-\alpha_{_1}-\alpha_{_2})y_{_1}\Big]
y_{_1}{\partial\over\partial y_{_1}}
\nonumber\\
&&\hspace{1.1cm}
-(1-\alpha_{_1}-\alpha_{_2})y_{_1}\Big(y_{_2}{\partial\over\partial y_{_2}}+y_{_3}{\partial\over\partial y_{_3}}\Big)
+(\alpha_{_3}\alpha_{_{6}}-\alpha_{_1}\alpha_{_2}y_{_1})
\;,\\
&&\hat{L}_{_2}=(1-y_{_2})y_{_2}^2{\partial^2\over\partial y_{_2}^2}-y_{_2}\Big(y_{_1}^2{\partial^2\over\partial y_{_1}^2}+y_{_3}^2{\partial^2\over\partial y_{_3}^2}\Big)
-y_{_2}\Big(y_{_1}y_{_2}{\partial^2\over\partial y_{_1}\partial y_{_2}}+y_{_1}y_{_3}{\partial^2\over\partial y_{_1}\partial y_{_3}}
\nonumber\\
&&\hspace{1.1cm}
+y_{_2}y_{_3}{\partial^2\over\partial y_{_2}\partial y_{_3}}\Big) +\Big[(1+\alpha_{_4}+\alpha_{_{7}})-(1-\alpha_{_1}-\alpha_{_2})y_{_2}\Big]
y_{_2}{\partial\over\partial y_{_2}}
\nonumber\\
&&\hspace{1.1cm}
-(1-\alpha_{_1}-\alpha_{_2})y_{_2}\Big(y_{_1}{\partial\over\partial y_{_1}}+y_{_3}{\partial\over\partial y_{_3}}\Big)
+(\alpha_{_4}\alpha_{_{7}}-\alpha_{_1}\alpha_{_2}y_{_2})
\;,\\
&&\hat{L}_{_3}=(1-y_{_3})y_{_3}^2{\partial^2\over\partial y_{_3}^2}-y_{_3}\Big(y_{_1}^2{\partial^2\over\partial y_{_1}^2}+y_{_2}^2{\partial^2\over\partial y_{_2}^2}\Big)
-y_{_3}\Big(y_{_1}y_{_2}{\partial^2\over\partial y_{_1}\partial y_{_2}}+y_{_1}y_{_3}{\partial^2\over\partial y_{_1}\partial y_{_3}}
\nonumber\\
&&\hspace{1.1cm}
+y_{_2}y_{_3}{\partial^2\over\partial y_{_2}\partial y_{_3}}\Big)
+\Big[(1+\alpha_{_5}+\alpha_{_{8}})-(1-\alpha_{_1}-\alpha_{_2})y_{_3}\Big]
y_{_3}{\partial\over\partial y_{_3}}
\nonumber\\
&&\hspace{1.1cm}
-(1-\alpha_{_1}-\alpha_{_2})y_{_3}\Big(y_{_1}{\partial\over\partial y_{_1}}+y_{_2}{\partial\over\partial y_{_2}}\Big)
+(\alpha_{_5}\alpha_{_{8}}-\alpha_{_1}\alpha_{_2}y_{_3})
\;.
\label{3loop-28}
\end{eqnarray}
The above system of linear PDEs in Eq. (\ref{3loop-27}) is too complicated to construct the solutions. In the following section, we will show the hypergeometric series solutions solving from the GKZ hypergeometric system in Eq. (\ref{3loop-20}).

\section{The hypergeometric series solutions of the three-loop vacuum integral with four propagates\label{sec3}}
\indent\indent

To construct the hypergeometric series solutions of the GKZ hypergeometric system of the three-loop vacuum integral with four propagates in Eq.~(\ref{3loop-20})
together with corresponding hyperbolic equations in Eq.~(\ref{3loop-23}) through triangulation
is equivalent to choose a set of the linear independent column vectors of the matrix in Eq.~(\ref{3loop-22})
which spans the dual space. We denote the submatrix
composed of the first, third, and fourth column vectors of the dual
matrix of Eq.~(\ref{3loop-22}) as $\mathbf{\tilde A}_{_{134}}$, i.e.
\begin{eqnarray}
&&\mathbf{\tilde A}_{_{134}}=\left(\begin{array}{ccc}
-1\;\;&1\;\;&0\;\;\\
-1\;\;&0\;\;&1\;\;\\
-1\;\;&0\;\;&0\;\;
\end{array}\right)\;.
\label{3loop-S1}
\end{eqnarray}
Obviously $\det\mathbf{\tilde A}_{_{134}}=-1\neq0$, and
\begin{eqnarray}
&&\mathbf{B}_{_{134}}=\mathbf{\tilde A}_{_{134}}^{-1}\cdot\mathbf{\tilde A}
\nonumber\\
&&\hspace{1.0cm}=\left(\begin{array}{cccccccc}
1\;\;& 1\;\;& 0\;\;& 0\;\;& -1\;\;& 0\;\;& 0\;\;& -1\;\;\\
0\;\;& 0\;\;& 1\;\;& 0\;\;& -1\;\;& 1\;\;& 0\;\;& -1\;\;\\
0\;\;& 0\;\;& 0\;\;& 1\;\;& -1\;\;& 0\;\;& 1\;\;& -1\;\;
\end{array}\right)\;.
\label{3loop-S2}
\end{eqnarray}
Taking 3 row vectors of the matrix $\mathbf{B}_{_{134}}$ as the basis of integer lattice,
one constructs the GKZ hypergeometric series solutions in parameter space through choosing the sets of column indices $I_{_i}\subset [1,8]\;(i=1,\cdots,8)$ which are consistent with the basis of integer lattice  $\mathbf{B}_{_{134}}$.

We take the set of column indices
$I_{_1}=[2,5,6,7,8]$, i.e. the implement $J_{_1}=[1,8]\setminus I_{_1}=[1,3,4]$.
The choice on the set of indices implies the exponent numbers
$\alpha_{_1}=\alpha_{_3}=\alpha_{_4}=0$. Through Eq.~(\ref{3loop-26-1}), one can have
\begin{eqnarray}
&&\alpha_{_2}=a_{_1}-a_{_2},\;\alpha_{_5}=b_{_1}+b_{_2}+b_{_3}-a_{_1}-3,\;
\nonumber\\
&&\alpha_{_6}=1-b_{_1},\;\alpha_{_7}=1-b_{_2},\;\alpha_{_8}=b_{_1}+b_{_2}-a_{_1}-2\;.
\label{3loop-S3}
\end{eqnarray}
Combined with Eq.~(\ref{3loop-11}), we can have
\begin{eqnarray}
\alpha_{_2}=\frac{D}{2}-1,\;\alpha_{_5}=-\frac{D}{2},\;
\alpha_{_6}=\frac{D}{2}-1,\;\alpha_{_7}=\frac{D}{2}-1,\;\alpha_{_8}=-1\;.
\label{3loop-S4}
\end{eqnarray}
According the basis of integer lattice $\mathbf{B}_{_{134}}$, the corresponding hypergeometric series solution with triple independent variables is written as
\begin{eqnarray}
&&\Phi_{_{[134]}}^{(1)}(\alpha,z)=\prod\limits_{i=1}^{8}z_{_i}^{\alpha_{_i}}\sum\limits_{n_{_1}=0}^\infty
\sum\limits_{n_{_2}=0}^\infty\sum\limits_{n_{_3}=0}^\infty
{c_{_{[134]}}^{(1)}(\alpha,{\bf n})}
\Big({z_{_1}z_{_2}\over z_{_5}z_{_{8}}}\Big)^{n_{_1}}
\Big({z_{_3}z_{_6}\over z_{_5}z_{_8}}\Big)^{n_{_2}}
\Big({z_{_4}z_{_7}\over z_{_5}z_{_{8}}}\Big)^{n_{_3}}
\nonumber\\
&&\hspace{2.0cm}=
\prod\limits_{i=1}^{8}z_{_i}^{\alpha_{_i}}\sum\limits_{n_{_1}=0}^\infty
\sum\limits_{n_{_2}=0}^\infty\sum\limits_{n_{_3}=0}^\infty
{c_{_{[134]}}^{(1)}(\alpha,{\bf n})}
\Big({1\over y_{_3}}\Big)^{n_{_1}}
\Big({y_{_1}\over y_{_3}}\Big)^{n_{_2}}
\Big({y_{_2}\over y_{_3}}\Big)^{n_{_3}}\;,
\label{3loop-S5}
\end{eqnarray}
with the coefficient is
\begin{eqnarray}
&&c_{_{[134]}}^{(1)}(\alpha,{\bf n})=\Big\{ n_{_1}!n_{_2}!n_{_3}!\Gamma(1+\alpha_{_2}+n_{_1})\Gamma(1+\alpha_{_5}-n_{_1}-n_{_2}-n_{_3})
\nonumber\\
&&\hspace{2.5cm}\times
\Gamma(1+\alpha_{_6}+n_{_2})\Gamma(1+\alpha_{_7}+n_{_3})
\Gamma(1+\alpha_{_8}-n_{_1}-n_{_2}-n_{_3})
\Big\}^{-1}\;.
\label{3loop-S6}
\end{eqnarray}
Using the relation in Eq.~(\ref{Gamma}),
one can have
\begin{eqnarray}
&&c_{_{[134]}}^{(1)}(\alpha,{\bf n})\propto
\frac{\Gamma(-\alpha_{_5}+n_{_1}+n_{_2}+n_{_3}) \Gamma(-\alpha_{_8}+n_{_1}+n_{_2}+n_{_3})}
{n_{_1}!n_{_2}!n_{_3}!\Gamma(1+\alpha_{_2}+n_{_1})\Gamma(1+\alpha_{_6}+n_{_2})\Gamma(1+\alpha_{_7}+n_{_3})}\;,
\label{3loop-S8}
\end{eqnarray}
where we ignore the constant term ${\frac{\sin\pi\alpha_{_5} \sin\pi\alpha_{_8}}{\pi^2}}$.
And then, through Eq.~(\ref{3loop-S4}), the corresponding hypergeometric series solution can be written as
\begin{eqnarray}
&&\Phi_{_{[134]}}^{(1)}(\alpha,z)=
y_{_1}^{{D\over2}-1}y_{_2}^{{D\over2}-1}y_{_3}^{-1}\sum\limits_{n_{_1}=0}^\infty
\sum\limits_{n_{_2}=0}^\infty\sum\limits_{n_{_3}=0}^\infty
{c_{_{[134]}}^{(1)}(\alpha,{\bf n})}
\Big({1\over y_{_3}}\Big)^{n_{_1}}
\Big({y_{_1}\over y_{_3}}\Big)^{n_{_2}}
\Big({y_{_2}\over y_{_3}}\Big)^{n_{_3}}\;,
\label{3loop-S9}
\end{eqnarray}
with the coefficient is
\begin{eqnarray}
&&c_{_{[134]}}^{(1)}(\alpha,{\bf n})=
\frac{\Gamma({D\over2}+n_{_1}+n_{_2}+n_{_3}) \Gamma(1+n_{_1}+n_{_2}+n_{_3})}
{n_{_1}!n_{_2}!n_{_3}!\Gamma({D\over2}+n_{_1})\Gamma({D\over2}+n_{_2})\Gamma({D\over2}+n_{_3})}\;.
\label{3loop-S10}
\end{eqnarray}
Here, the convergent region
of the hypergeometric function $\Phi_{_{[134]}}^{(1)}(\alpha,z)$ in Eq.~(\ref{3loop-S9}) is
\begin{eqnarray}
&&\Xi_{_{[134]}}=\{(y_{_1},\;y_{_2},\;y_{_3})
\Big|1<|y_{_3}|,\;|y_{_1}|<|y_{_3}|,\;|y_{_2}|<|y_{_3}|\}\;,
\label{3loop-S11}
\end{eqnarray}
which shows that $\Phi_{_{[134]}}^{(1)}(\alpha,z)$ is in neighborhood of regular singularity $\infty$.

In a similar way, we can obtain other seven hypergeometric solutions which
are consistent with the basis of integer lattice $\mathbf{B}_{_{134}}$, and the convergent region is also $\Xi_{_{[134]}}$.
\begin{itemize}
\item $I_{_2}=[2,4,5,6,8]$, i.e.
the implement $J_{_2}=[1,8]\setminus I_{_2}=[1,3,7]$.
The choice implies the exponent numbers $\alpha_{_1}=\alpha_{_3}=\alpha_{_7}=0$, and
\begin{eqnarray}
\alpha_{_2}=\alpha_{_6}=\frac{D}{2}-1,\;\alpha_{_4}=1-\frac{D}{2},\;\alpha_{_5}=-1,\;
\alpha_{_8}=\frac{D}{2}-2\;.
\label{3loop-S12-1}
\end{eqnarray}
The corresponding hypergeometric series solution is written as
\begin{eqnarray}
&&\Phi_{_{[134]}}^{(2)}(\alpha,z)=
y_{_1}^{{D\over2}-1}y_{_3}^{{D\over2}-2}\sum\limits_{n_{_1}=0}^\infty
\sum\limits_{n_{_2}=0}^\infty\sum\limits_{n_{_3}=0}^\infty
{c_{_{[134]}}^{(2)}(\alpha,{\bf n})}
\Big({1\over y_{_3}}\Big)^{n_{_1}}
\Big({y_{_1}\over y_{_3}}\Big)^{n_{_2}}
\Big({y_{_2}\over y_{_3}}\Big)^{n_{_3}}\;,
\label{3loop-S12-2}
\end{eqnarray}
with the coefficient is
\begin{eqnarray}
&&c_{_{[134]}}^{(2)}(\alpha,{\bf n})=
\frac{\Gamma(2-{D\over2}+n_{_1}+n_{_2}+n_{_3}) \Gamma(1+n_{_1}+n_{_2}+n_{_3})}
{n_{_1}!n_{_2}!n_{_3}!\Gamma({D\over2}+n_{_1})\Gamma({D\over2}+n_{_2})\Gamma(2-{D\over2}+n_{_3})}\;.
\label{3loop-S12-3}
\end{eqnarray}

\item $I_{_3}=[2,3,5,7,8]$, i.e.
the implement $J_{_3}=[1,8]\setminus I_{_3}=[1,4,6]$.
The choice implies the exponent numbers $\alpha_{_1}=\alpha_{_4}=\alpha_{_6}=0$, and
\begin{eqnarray}
\alpha_{_2}=\alpha_{_7}=\frac{D}{2}-1,\;\alpha_{_3}=1-\frac{D}{2},\;\alpha_{_5}=-1,\;
\alpha_{_8}=\frac{D}{2}-2\;.
\label{3loop-S13-1}
\end{eqnarray}
The corresponding hypergeometric series solution is written as
\begin{eqnarray}
&&\Phi_{_{[134]}}^{(3)}(\alpha,z)=
y_{_2}^{{D\over2}-1}y_{_3}^{{D\over2}-2}\sum\limits_{n_{_1}=0}^\infty
\sum\limits_{n_{_2}=0}^\infty\sum\limits_{n_{_3}=0}^\infty
{c_{_{[134]}}^{(3)}(\alpha,{\bf n})}
\Big({1\over y_{_3}}\Big)^{n_{_1}}
\Big({y_{_1}\over y_{_3}}\Big)^{n_{_2}}
\Big({y_{_2}\over y_{_3}}\Big)^{n_{_3}}\;,
\label{3loop-S13-2}
\end{eqnarray}
with the coefficient is
\begin{eqnarray}
&&c_{_{[134]}}^{(3)}(\alpha,{\bf n})=
\frac{\Gamma(2-{D\over2}+n_{_1}+n_{_2}+n_{_3}) \Gamma(1+n_{_1}+n_{_2}+n_{_3})}
{n_{_1}!n_{_2}!n_{_3}!\Gamma({D\over2}+n_{_1})\Gamma(2-{D\over2}+n_{_2})\Gamma({D\over2}+n_{_3})}\;.
\label{3loop-S13-3}
\end{eqnarray}

\item $I_{_4}=[2,3,4,5,8]$, i.e.
the implement $J_{_4}=[1,8]\setminus I_{_4}=[1,6,7]$.
The choice implies the exponent numbers $\alpha_{_1}=\alpha_{_6}=\alpha_{_7}=0$, and
\begin{eqnarray}
\alpha_{_2}=\frac{D}{2}-1,\;\alpha_{_3}=\alpha_{_4}=1-\frac{D}{2},\;
\alpha_{_5}=\frac{D}{2}-2,\;\alpha_{_8}=D-3\;.
\label{3loop-S14-1}
\end{eqnarray}
The corresponding hypergeometric series solution is written as
\begin{eqnarray}
&&\Phi_{_{[134]}}^{(4)}(\alpha,z)=
y_{_3}^{D-3}\sum\limits_{n_{_1}=0}^\infty
\sum\limits_{n_{_2}=0}^\infty\sum\limits_{n_{_3}=0}^\infty
{c_{_{[134]}}^{(4)}(\alpha,{\bf n})}
\Big({1\over y_{_3}}\Big)^{n_{_1}}
\Big({y_{_1}\over y_{_3}}\Big)^{n_{_2}}
\Big({y_{_2}\over y_{_3}}\Big)^{n_{_3}}\;,
\label{3loop-S14-2}
\end{eqnarray}
with the coefficient is
\begin{eqnarray}
&&c_{_{[134]}}^{(4)}(\alpha,{\bf n})=
\frac{\Gamma(2-{D\over2}+n_{_1}+n_{_2}+n_{_3}) \Gamma(3-D+n_{_1}+n_{_2}+n_{_3})}
{n_{_1}!n_{_2}!n_{_3}!\Gamma({D\over2}+n_{_1})\Gamma(2-{D\over2}+n_{_2})\Gamma(2-{D\over2}+n_{_3})}\;.
\label{3loop-S14-3}
\end{eqnarray}

\item $I_{_5}=[1,5,6,7,8]$, i.e.
the implement $J_{_5}=[1,8]\setminus I_{_5}=[2,3,4]$.
The choice implies the exponent numbers $\alpha_{_2}=\alpha_{_3}=\alpha_{_4}=0$, and
\begin{eqnarray}
\alpha_{_1}=1-\frac{D}{2},\;\alpha_{_5}=-1,\;
\alpha_{_6}=\alpha_{_7}=\frac{D}{2}-1,\;\alpha_{_8}=\frac{D}{2}-2\;.
\label{3loop-S15-1}
\end{eqnarray}
The corresponding hypergeometric series solution is written as
\begin{eqnarray}
&&\Phi_{_{[134]}}^{(5)}(\alpha,z)=
y_{_1}^{{D\over2}-1}y_{_2}^{{D\over2}-1}y_{_3}^{{D\over2}-2}\sum\limits_{n_{_1}=0}^\infty
\sum\limits_{n_{_2}=0}^\infty\sum\limits_{n_{_3}=0}^\infty
{c_{_{[134]}}^{(5)}(\alpha,{\bf n})}
\Big({1\over y_{_3}}\Big)^{n_{_1}}
\Big({y_{_1}\over y_{_3}}\Big)^{n_{_2}}
\Big({y_{_2}\over y_{_3}}\Big)^{n_{_3}}\;,
\label{3loop-S15-2}
\end{eqnarray}
with the coefficient is
\begin{eqnarray}
&&c_{_{[134]}}^{(5)}(\alpha,{\bf n})=
\frac{\Gamma(2-{D\over2}+n_{_1}+n_{_2}+n_{_3}) \Gamma(1+n_{_1}+n_{_2}+n_{_3})}
{n_{_1}!n_{_2}!n_{_3}!\Gamma(2-{D\over2}+n_{_1})\Gamma({D\over2}+n_{_2})\Gamma({D\over2}+n_{_3})}\;.
\label{3loop-S15-3}
\end{eqnarray}

\item $I_{_6}=[1,4,5,6,8]$, i.e.
the implement $J_{_6}=[1,8]\setminus I_{_6}=[2,3,7]$.
The choice implies the exponent numbers $\alpha_{_2}=\alpha_{_3}=\alpha_{_7}=0$, and
\begin{eqnarray}
\alpha_{_1}=\alpha_{_4}=1-\frac{D}{2},\;\alpha_{_5}=\frac{D}{2}-2,\;
\alpha_{_6}=\frac{D}{2}-1,\;\alpha_{_8}=D-3\;.
\label{3loop-S16-1}
\end{eqnarray}
The corresponding hypergeometric series solution is written as
\begin{eqnarray}
&&\Phi_{_{[134]}}^{(6)}(\alpha,z)=
y_{_1}^{{D\over2}-1}y_{_3}^{D-3}\sum\limits_{n_{_1}=0}^\infty
\sum\limits_{n_{_2}=0}^\infty\sum\limits_{n_{_3}=0}^\infty
{c_{_{[134]}}^{(6)}(\alpha,{\bf n})}
\Big({1\over y_{_3}}\Big)^{n_{_1}}
\Big({y_{_1}\over y_{_3}}\Big)^{n_{_2}}
\Big({y_{_2}\over y_{_3}}\Big)^{n_{_3}}\;,
\label{3loop-S16-2}
\end{eqnarray}
with the coefficient is
\begin{eqnarray}
&&c_{_{[134]}}^{(6)}(\alpha,{\bf n})=
\frac{\Gamma(2-{D\over2}+n_{_1}+n_{_2}+n_{_3}) \Gamma(3-D+n_{_1}+n_{_2}+n_{_3})}
{n_{_1}!n_{_2}!n_{_3}!\Gamma(2-{D\over2}+n_{_1})\Gamma({D\over2}+n_{_2})\Gamma(2-{D\over2}+n_{_3})}\;.
\label{3loop-S16-3}
\end{eqnarray}

\item $I_{_7}=[1,3,5,7,8]$, i.e.
the implement $J_{_7}=[1,8]\setminus I_{_7}=[2,4,6]$.
The choice implies the exponent numbers $\alpha_{_2}=\alpha_{_4}=\alpha_{_6}=0$, and
\begin{eqnarray}
\alpha_{_1}=\alpha_{_3}=1-\frac{D}{2},\;\alpha_{_5}=\frac{D}{2}-2,\;
\alpha_{_7}=\frac{D}{2}-1,\;\alpha_{_8}=D-3\;.
\label{3loop-S17-1}
\end{eqnarray}
The corresponding hypergeometric series solution is written as
\begin{eqnarray}
&&\Phi_{_{[134]}}^{(7)}(\alpha,z)=
y_{_2}^{{D\over2}-1}y_{_3}^{D-3}\sum\limits_{n_{_1}=0}^\infty
\sum\limits_{n_{_2}=0}^\infty\sum\limits_{n_{_3}=0}^\infty
{c_{_{[134]}}^{(7)}(\alpha,{\bf n})}
\Big({1\over y_{_3}}\Big)^{n_{_1}}
\Big({y_{_1}\over y_{_3}}\Big)^{n_{_2}}
\Big({y_{_2}\over y_{_3}}\Big)^{n_{_3}}\;,
\label{3loop-S17-2}
\end{eqnarray}
with the coefficient is
\begin{eqnarray}
&&c_{_{[134]}}^{(7)}(\alpha,{\bf n})=
\frac{\Gamma(2-{D\over2}+n_{_1}+n_{_2}+n_{_3}) \Gamma(3-D+n_{_1}+n_{_2}+n_{_3})}
{n_{_1}!n_{_2}!n_{_3}!\Gamma(2-{D\over2}+n_{_1})\Gamma(2-{D\over2}+n_{_2})\Gamma({D\over2}+n_{_3})}\;.
\label{3loop-S17-3}
\end{eqnarray}

\item $I_{_8}=[1,3,4,5,8]$, i.e.
the implement $J_{_8}=[1,8]\setminus I_{_8}=[2,6,7]$.
The choice implies the exponent numbers $\alpha_{_2}=\alpha_{_6}=\alpha_{_7}=0$, and
\begin{eqnarray}
\alpha_{_1}=1-\frac{D}{2},\;\alpha_{_3}=\alpha_{_4}=1-\frac{D}{2},\;
\alpha_{_5}=D-3,\;\alpha_{_8}=\frac{3D}{2}-4\;.
\label{3loop-S18-1}
\end{eqnarray}
The corresponding hypergeometric series solution is written as
\begin{eqnarray}
&&\Phi_{_{[134]}}^{(8)}(\alpha,z)=
y_{_3}^{{3D\over2}-4}\sum\limits_{n_{_1}=0}^\infty
\sum\limits_{n_{_2}=0}^\infty\sum\limits_{n_{_3}=0}^\infty
{c_{_{[134]}}^{(8)}(\alpha,{\bf n})}
\Big({1\over y_{_3}}\Big)^{n_{_1}}
\Big({y_{_1}\over y_{_3}}\Big)^{n_{_2}}
\Big({y_{_2}\over y_{_3}}\Big)^{n_{_3}}\;,
\label{3loop-S18-2}
\end{eqnarray}
with the coefficient is
\begin{eqnarray}
&&c_{_{[134]}}^{(8)}(\alpha,{\bf n})=
\frac{\Gamma(4-{3D\over2}+n_{_1}+n_{_2}+n_{_3}) \Gamma(3-D+n_{_1}+n_{_2}+n_{_3})}
{n_{_1}!n_{_2}!n_{_3}!\Gamma(2-{D\over2}+n_{_1})\Gamma(2-{D\over2}+n_{_2})\Gamma(2-{D\over2}+n_{_3})}\;.
\label{3loop-S18-3}
\end{eqnarray}

\end{itemize}
The above eight hypergeometric series solutions $\Phi_{_{[134]}}^{(i)}(\alpha,z)$ whose convergent region is  $\Xi_{_{[134]}}$ can constitute a fundamental solution system.  The combination coefficients are determined by the value of the scalar integral of an ordinary point or some regular singularities.

Multiplying one of the row vectors of the matrix $\mathbf{B}_{_{134}}$ by -1,
the induced integer matrix can also be chosen as a basis of the integer lattice
space of certain hypergeometric series. Taking 3 row vectors of the following matrix as the
basis of integer lattice,
\begin{eqnarray}
&&\mathbf{B}_{_{\tilde{1}34}}={\rm diag}(-1,1,1)\cdot\mathbf{B}_{_{134}}
\nonumber\\
&&\hspace{0.9cm}=\left(\begin{array}{cccccccc}
-1\;\;& -1\;\;& 0\;\;& 0\;\;& 1\;\;& 0\;\;& 0\;\;& 1\;\;\\
0\;\;& 0\;\;& 1\;\;& 0\;\;& -1\;\;& 1\;\;& 0\;\;& -1\;\;\\
0\;\;& 0\;\;& 0\;\;& 1\;\;& -1\;\;& 0\;\;& 1\;\;& -1\;\;
\end{array}\right)\;,
\label{3loop-S19}
\end{eqnarray}
one obtains eight hypergeometric series solutions  $\Phi_{_{[\tilde{1}34]}}^{(i)}(\alpha,z)\:(i=1,\cdots,8)$  similarly, which the expressions are collected in Appendix~\ref{app1}.
The convergent region of the hypergeometric functions $\Phi_{_{[\tilde{1}34]}}^{(i)}(\alpha,z)$ is
\begin{eqnarray}
&&\Xi_{_{[\tilde{1}34]}}=\{(y_{_1},\;y_{_2},\;y_{_3})
\Big||y_{_1}|<1,\;|y_{_2}|<1,\;|y_{_3}|<1\}\;,
\label{3loop-S20}
\end{eqnarray}
which shows that $\Phi_{_{[\tilde{1}34]}}^{(i)}(\alpha,z)$ are in neighborhood of regular singularity $0$ and  can constitute a fundamental solution system.

Taking 3 row vectors of the following matrix as the basis of integer lattice,
\begin{eqnarray}
&&\mathbf{B}_{_{1\tilde{3}4}}={\rm diag}(1,-1,1)\cdot\mathbf{B}_{_{134}}
\nonumber\\
&&\hspace{0.9cm}=\left(\begin{array}{cccccccc}
1\;\;& 1\;\;& 0\;\;& 0\;\;& -1\;\;& 0\;\;& 0\;\;& -1\;\;\\
0\;\;& 0\;\;& -1\;\;& 0\;\;& 1\;\;& -1\;\;& 0\;\;& 1\;\;\\
0\;\;& 0\;\;& 0\;\;& 1\;\;& -1\;\;& 0\;\;& 1\;\;& -1\;\;
\end{array}\right)\;,
\label{3loop-S39}
\end{eqnarray}
one obtains eight hypergeometric series solutions  $\Phi_{_{[1\tilde{3}4]}}^{(i)}(\alpha,z)\:(i=1,\cdots,8)$  similarly, which the expressions are collected in Appendix~\ref{app2}.
The convergent region of the hypergeometric functions $\Phi_{_{[1\tilde{3}4]}}^{(i)}(\alpha,z)$ is
\begin{eqnarray}
&&\Xi_{_{[1\tilde{3}4]}}=\{(y_{_1},\;y_{_2},\;y_{_3})
\Big|1<|y_{_1}|,\;|y_{_2}|<|y_{_1}|,\;|y_{_3}|<|y_{_1}|\}\;,
\label{3loop-S40}
\end{eqnarray}
which shows that $\Phi_{_{[1\tilde{3}4]}}^{(i)}(\alpha,z)$ are in neighborhood of regular singularity $\infty$ and can constitute a fundamental solution system.

Taking 3 row vectors of the following matrix as the basis of integer lattice,
\begin{eqnarray}
&&\mathbf{B}_{_{13\tilde{4}}}={\rm diag}(1,1,-1)\cdot\mathbf{B}_{_{134}}
\nonumber\\
&&\hspace{0.9cm}=\left(\begin{array}{cccccccc}
1\;\;& 1\;\;& 0\;\;& 0\;\;& -1\;\;& 0\;\;& 0\;\;& -1\;\;\\
0\;\;& 0\;\;& 1\;\;& 0\;\;& -1\;\;& 1\;\;& 0\;\;& -1\;\;\\
0\;\;& 0\;\;& 0\;\;& -1\;\;& 1\;\;& 0\;\;& -1\;\;& 1\;\;
\end{array}\right)\;,
\label{3loop-S49}
\end{eqnarray}
one obtains eight hypergeometric series solutions  $\Phi_{_{[13\tilde{4}]}}^{(i)}(\alpha,z)\:(i=1,\cdots,8)$  similarly, which the expressions are collected in Appendix~\ref{app3}.
The convergent region of the hypergeometric functions $\Phi_{_{[13\tilde{4}]}}^{(i)}(\alpha,z)$ is
\begin{eqnarray}
&&\Xi_{_{[1\tilde{3}4]}}=\{(y_{_1},\;y_{_2},\;y_{_3})
\Big|1<|y_{_2}|,\;|y_{_1}|<|y_{_2}|,\;|y_{_3}|<|y_{_2}|\}\;,
\label{3loop-S50}
\end{eqnarray}
which shows that $\Phi_{_{[13\tilde{4}]}}^{(i)}(\alpha,z)$ are in neighborhood of regular singularity $\infty$ and can constitute a fundamental solution system.

The above hypergeometric series solutions are consistent with our previous work for the three-loop vacuum integral with four propagates \cite{{Gu2019}}, which  the  hypergeometric functions of the scalar integral can be obtained using the power series of modified Bessel functions and the radial integral. And some special cases  for the three-loop vacuum integral with four propagates are also shown there.

\section{GKZ hypergeometric system of the three-loop vacuum integral with five propagates\label{sec-five}}
\indent\indent

\begin{figure}[ht]
\setlength{\unitlength}{0cm}
\centering
\vspace{0.0cm}\hspace{2cm}
\includegraphics[width=8.0cm]{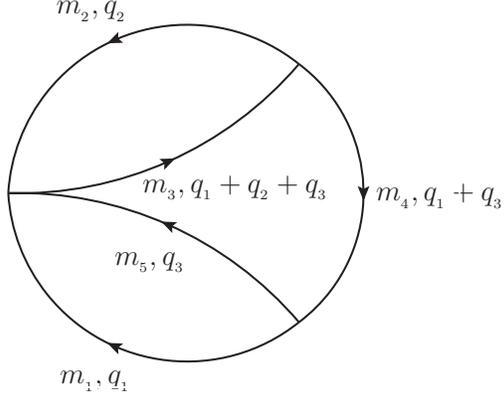}
\vspace{0cm}
\caption[]{Three-loop vacuum diagram with five propagators, which $m_{_i}$ denotes the mass of the $i$-th particle and $q_{_j}$ denotes the momentum.}
\label{fig-3loop5}
\end{figure}

The general analytic expression for the Feynman integral of a three-loop vacuum diagram with five propagates in Fig.~\ref{fig-3loop5} is written as
\begin{eqnarray}
&&U_{_5}=\Big(\Lambda_{_{\rm RE}}^2\Big)^{6-\frac{3D}{2}}\int{d^Dq_{_1}\over(2\pi)^D}
{d^Dq_{_2}\over(2\pi)^D}{d^Dq_{_3}\over(2\pi)^D}
\nonumber\\
&&\hspace{1.1cm}\times
{1\over(q_{_1}^2-m_{_1}^2)(q_{_2}^2-m_{_2}^2)((q_{_1}+q_{_2}+q_{_3})^2-m_{_3}^2)
((q_{_1}+q_{_3})^2-m_{_4}^2)(q_{_3}^2-m_{_5}^2)},
\label{GKZ0}
\end{eqnarray}
Adopting the notation of Refs.~\cite{Feng2018,Feng2019}, the Feynman integral of the three-loop vacuum diagram with five propagates can be written as
\begin{eqnarray}
&&U_{_5}=\frac{\Big(\Lambda_{_{\rm RE}}^2\Big)^{6-\frac{3D}{2}}}{(2\pi i)^4} \int_{-i\infty}^{+i\infty}ds_{_1}ds_{_2}ds_{_3}ds_{_4}
\Big[\prod\limits_{i=1}^4(-m_{_i}^2)^{s_{_i}}\Gamma(-s_{_i})\Gamma(1+s_{_i})\Big]
I_{Q}\;,
\label{GKZ1}
\end{eqnarray}
where
\begin{eqnarray}
I_{Q}\equiv \int{d^Dq_{_1}\over(2\pi)^D}{d^Dq_{_2}\over(2\pi)^D} {d^Dq_{_3}\over(2\pi)^D}
{1\over(q_{_1}^2)^{1+s_{_1}}(q_{_2}^2)^{1+s_{_2}}((q_{_1}+q_{_2}+q_{_3})^2)^{1+s_{_3}}
((q_{_1}+q_{_3})^2)^{1+s_{_4}}(q_{_3}^2-m_{_5}^2)}.\;\;
\label{GKZ1-0}
\end{eqnarray}

First, we can integrate out $q_{_2}$:
\begin{eqnarray}
I_{Q}&&=
{i(-)^{D/2}\Gamma(2-{D\over2}+s_{_2}+s_{_3})\Gamma({D\over2}-1-s_{_2})\Gamma({D\over2}-1-s_{_3})
\over(4\pi)^{D/2}\Gamma(1+s_{_2})\Gamma(1+s_{_3})\Gamma(D-2-s_{_2}-s_{_3})}
\nonumber\\
&&\hspace{0.4cm}\times
\int{d^Dq_{_1}\over(2\pi)^D}{d^Dq_{_3}\over(2\pi)^D}
{1\over(q_{_1}^2)^{1+s_{_1}}((q_{_1}+q_{_3})^2)^{3-{D\over2}+s_{_2}+s_{_3}+s_{_4}}(q_{_3}^2-m_{_5}^2)}.
\label{GKZ1-1}
\end{eqnarray}
Second, one can integrate out $q_{_1}$:
\begin{eqnarray}
&&\int{d^Dq_{_1}\over(2\pi)^D}{d^Dq_{_3}\over(2\pi)^D}
{1\over(q_{_1}^2)^{1+s_{_1}}((q_{_1}+q_{_3})^2)^{3-{D\over2}+s_{_2}+s_{_3}+s_{_4}}(q_{_3}^2-m_{_5}^2)}
\nonumber\\
&&\hspace{-0.0cm}=
{i(-)^{D/2}\Gamma(4-D+\sum\limits_{i=1}^4s_{_i})\Gamma({D\over2}-1-s_{_1})\Gamma(D-3-\sum\limits_{i=2}^4s_{_i})
\over(4\pi)^{D/2}\Gamma(1+s_{_1})\Gamma(3-{D\over2}+\sum\limits_{i=2}^4s_{_i})
\Gamma({3D\over2}-4-\sum\limits_{i=1}^4s_{_i})}
\nonumber\\
&&\hspace{0.4cm}\times
\int{d^Dq_{_3}\over(2\pi)^D}
{1\over(q_{_3}^2-m_{_5}^2)(q_{_3}^2)^{4-D+s_{_1}+s_{_2}+s_{_3}+s_{_4}}}.
\label{GKZ1-2}
\end{eqnarray}
Third, we integrate out $q_{_3}$:
\begin{eqnarray}
&&\int{d^Dq_{_3}\over(2\pi)^D}
{1\over(q_{_3}^2-m_{_5}^2)(q_{_3}^2)^{4-D+s_{_1}+s_{_2}+s_{_3}+s_{_4}}}
\nonumber\\
&&\hspace{-0.0cm}=
{{i} \over (4\pi)^{D/2}} (-)^{5-D+\sum\limits_{i=1}^4s_{_i}} \Big( {1\over m_{_5}^2}\Big)^{5-{3D\over2}+\sum\limits_{i=1}^4s_{_i}}
{\Gamma(5-{3D\over2}+\sum\limits_{i=1}^4s_{_i}) \Gamma({3D\over2}-4-\sum\limits_{i=1}^4s_{_i}) }.
\label{GKZ1-3}
\end{eqnarray}

Together with Eqs. (\ref{GKZ1-1}-\ref{GKZ1-3}), one can have
\begin{eqnarray}
I_{Q}&&={{i} \over (4\pi)^{3D\over2}} (-)^{\sum\limits_{i=1}^4s_{_i}} \Big( {1\over m_{_5}^2}\Big)^{5-{3D\over2}+\sum\limits_{i=1}^4s_{_i}}
\Big[\prod\limits_{i=1}^3\Gamma({D\over2}-1-s_{_i})\Gamma(1+s_{_i})^{-1}\Big]\Gamma(4-D+\sum\limits_{i=1}^4s_{_i})
\nonumber\\
&&\hspace{0.4cm}\times
{\Gamma(2-{D\over2}+s_{_2}+s_{_3})\Gamma(D-3-\sum\limits_{i=2}^4s_{_i})
\Gamma(5-{3D\over2}+\sum\limits_{i=1}^4s_{_i})
\over\Gamma(D-2-s_{_2}-s_{_3})\Gamma(3-{D\over2}+\sum\limits_{i=2}^4s_{_i})}\;.
\label{GKZ2-1}
\end{eqnarray}
Then, the Mellin-Barnes representation of the Feynman integral of the three-loop vacuum diagram  with five propagates can be written as
\begin{eqnarray}
&&U_{_5}=
{im_{_5}^2\over(2\pi i)^4(4\pi)^6}\Big({4\pi\Lambda_{_{\rm RE}}^2\over m_{_5}^2}\Big)^{6-{3D\over2}}
\int_{-i\infty}^{+i\infty}ds_{_1}ds_{_2}ds_{_3}ds_{_4}
\nonumber\\
&&\hspace{1.0cm}\times
\Big[\prod\limits_{i=1}^4\Big({m_{_i}^2\over m_{_5}^2}\Big)^{s_{_i}}\Gamma(-s_{_i})\Big]\Gamma(1+s_{_4})
\Big[\prod\limits_{i=1}^3\Gamma({D\over2}-1-s_{_i})\Big]\Gamma(4-D+\sum\limits_{i=1}^4s_{_i})
\nonumber\\
&&\hspace{1.0cm}\times
{\Gamma(2-{D\over2}+s_{_2}+s_{_3})\Gamma(D-3-\sum\limits_{i=2}^4s_{_i})
\Gamma(5-{3D\over2}+\sum\limits_{i=1}^4s_{_i})
\over\Gamma(D-2-s_{_2}-s_{_3})\Gamma(3-{D\over2}+\sum\limits_{i=2}^4s_{_i})}\;.
\label{GKZ2}
\end{eqnarray}

Taking the residue of the pole of $\Gamma(-s_{_i}),\;(i=1,\;2,\;3,\;4)$, one can derive one linear independent term of the vacuum integral:
\begin{eqnarray}
&&U_{_5}\ni
{im_{_5}^2\over(4\pi)^6}\Big({4\pi\Lambda_{_{\rm RE}}^2\over m_{_5}^2}\Big)^{6-{3D\over2}}
\sum\limits_{n_{_1}=0}^\infty\sum_{n_{_2}=0}^\infty\sum_{n_{_3}=0}^\infty\sum_{n_{_4}=0}^\infty
(-)^{\sum\limits_{i=1}^4n_{_i}}x_{_1}^{n_{_1}}x_{_2}^{n_{_2}}x_{_3}^{n_{_3}}x_{_4}^{n_{_4}}
\nonumber\\
&&\hspace{1.0cm}\times
{\Gamma(1+n_{_4})
\Big[\prod\limits_{i=1}^3\Gamma({D\over2}-1-n_{_i})\Big]\Gamma(4-D+\sum\limits_{i=1}^4n_{_i}) \over n_{_1}!n_{_2}!n_{_3}!n_{_4}!}
\nonumber\\
&&\hspace{1.0cm}\times
{\Gamma(2-{D\over2}+n_{_2}+n_{_3})\Gamma(D-3-\sum\limits_{i=2}^4n_{_i})
\Gamma(5-{3D\over2}+\sum\limits_{i=1}^4n_{_i})
\over\Gamma(D-2-n_{_2}-n_{_3})\Gamma(3-{D\over2}+\sum\limits_{i=2}^4n_{_i})},
\label{GKZ3}
\end{eqnarray}
with $x_{_i}={m_{_i}^2\over m_{_5}^2},\;(i=1,\;2,\;3,\;4)$.
Adopting the identity in Eq.~(\ref{Gamma}), Eq.~(\ref{GKZ3}) can be written as
\begin{eqnarray}
&&U_{_5}\ni
{im_{_5}^2\over(4\pi)^6}\Big({4\pi\Lambda_{_{\rm RE}}^2\over m_{_5}^2}\Big)^{6-{3D\over2}}
{ \pi^3\over \sin^3\frac{\pi D}{2} }
T_{_{5}}({\bf a},\;{\bf b}\;\Big|\;{\bf x})\;,
\label{GKZ5}
\end{eqnarray}
with
\begin{eqnarray}
T_{_{5}}({\bf a},\;{\bf b}\;\Big|\;{\bf x})=\sum\limits_{n_{_1}=0}^\infty\sum_{n_{_2}=0}^\infty\sum_{n_{_3}=0}^\infty\sum_{n_{_4}=0}^\infty
A_{_{n_{_1}n_{_2}n_{_3}n_{_4}}} x_{_1}^{n_{_1}}x_{_2}^{n_{_2}}x_{_3}^{n_{_3}}x_{_4}^{n_{_4}},
\label{GKZ6}
\end{eqnarray}
where ${\bf x}=(x_{_1},\;\cdots,\;x_{_4})$, ${\bf a}=(a_{_1},\;\cdots,a_{_5})$ and
${\bf b}=(b_{_1},\;\cdots,b_{_5})$ with
\begin{eqnarray}
&&a_{_1}=4-D,\;a_{_2}=5-{3D\over2},\;a_{_3}=2-{D\over2},\;
a_{_4}=3-D,\;a_{_5}=1,
\nonumber\\
&&b_{_1}=b_{_2}=b_{_3}=2-{D\over2},\;b_{_4}=3-{D\over2},\;b_{_5}=4-D\:,
\label{GKZ6-1}
\end{eqnarray}
and the coefficient $A_{_{n_{_1}n_{_2}n_{_3}n_{_4}}}$ is
\begin{eqnarray}
\hspace{-0.5cm}A_{_{n_{_1}n_{_2}n_{_3}n_{_4}}}=
{\Gamma(a_{_1}+\sum\limits_{i=1}^4n_{_i}) \Gamma(a_{_2}+\sum\limits_{i=1}^4n_{_i})\Gamma(a_{_3}+n_{_2}+n_{_3})\Gamma(a_{_4}+n_{_2}+n_{_3})\Gamma(a_{_5}+n_{_4})
\over  n_{_1}!n_{_2}!n_{_3}!n_{_4}!\Gamma(b_{_1}+n_{_i})\Gamma(b_{_2}+n_{_2})\Gamma(b_{_3}+n_{_3})
\Gamma(b_{_4}+\sum\limits_{i=2}^4n_{_i})\Gamma(b_{_5}+\sum\limits_{i=2}^4n_{_i})}.
\label{GKZ6-2}
\end{eqnarray}

In order to proceed with our analysis, we define the auxiliary function
\begin{eqnarray}
&&\Phi_{_{5}}({\bf a},\;{\bf b}\;\Big|\;{\bf x},\;{\bf u},\;{\bf v})={\bf u}^{\bf a}{\bf v}^{{\bf b}-{\bf e}}
T_{_{5}}({\bf a},\;{\bf b}\;\Big|\;{\bf x})\;,
\label{GKZ10}
\end{eqnarray}
with the intermediate variables ${\bf u}={\bf v}={\bf e}=(1,\;1,\;1,\;1,\;1)$.
Then one can obtain
\begin{eqnarray}
&&\vartheta_{_{u_j}}\Phi_{_{5}}({\bf a},\;{\bf b}\;\Big|\;{\bf x},\;{\bf u},\;{\bf v})=a_{_j}
\Phi_{_{5}}({\bf a},\;{\bf b}\;\Big|\;{\bf x},\;{\bf u},\;{\bf v})
\;,\nonumber\\
&&\vartheta_{_{v_j}}\Phi_{_{5}}({\bf a},\;{\bf b}\;\Big|\;{\bf x},\;{\bf u},\;{\bf v})=(b_{_j}-1)
\Phi_{_{5}}({\bf a},\;{\bf b}\;\Big|\;{\bf x},\;{\bf u},\;{\bf v})\;,
\label{GKZ11}
\end{eqnarray}
where $\vartheta_{_{u_{_j}}}=x_{_j}\partial_{_{u_{_j}}}$
denotes the Euler operators, and $\partial_{_{u_{_j}}}=\partial/\partial u_{_j}$, respectively.

Through the transformation
\begin{eqnarray}
&&z_{_j}={1\over u_{_j}},\;\;z_{_{5+j}}=v_{_j},\;\;(j=1,\cdots,5),
\nonumber\\
&&z_{_{11}}={x_{_1}\over u_{_1}u_{_2}v_{_1}},\;\;z_{_{12}}={x_{_2}\over u_{_1}u_{_2}u_{_3}u_{_4}v_{_2}v_{_4}v_{_5}},
\nonumber\\
&&z_{_{13}}={x_{_3}\over u_{_1}u_{_2}u_{_3}u_{_4}v_{_3}v_{_4}v_{_5}},
\;\;z_{_{14}}={x_{_4}\over u_{_1}u_{_2}u_{_5}v_{_4}v_{_5}}\;,
\label{GKZ14}
\end{eqnarray}
one derives the GKZ hypergeometric system for the three-loop vacuum integral  with five propagates
\begin{eqnarray}
&&\mathbf{A_{_{5}}}\cdot\vec{\vartheta_{_{5}}}\Phi_{_{5}}=\mathbf{B_{_{5}}}\Phi_{_{5}}\;,
\label{GKZ17}
\end{eqnarray}
where
\begin{eqnarray}
&&\mathbf{A_{_{5}}}=\left(\begin{array}{cccccccccccccc}
1\;\;&0\;\;&0\;\;&0\;\;&0\;\;&0\;\;&0\;\;&0\;\;&0\;\;&0\;\;&1\;\;&1\;\;&1\;\;&1\;\;\\
0\;\;&1\;\;&0\;\;&0\;\;&0\;\;&0\;\;&0\;\;&0\;\;&0\;\;&0\;\;&1\;\;&1\;\;&1\;\;&1\;\;\\
0\;\;&0\;\;&1\;\;&0\;\;&0\;\;&0\;\;&0\;\;&0\;\;&0\;\;&0\;\;&0\;\;&1\;\;&1\;\;&0\;\;\\
0\;\;&0\;\;&0\;\;&1\;\;&0\;\;&0\;\;&0\;\;&0\;\;&0\;\;&0\;\;&0\;\;&1\;\;&1\;\;&0\;\;\\
0\;\;&0\;\;&0\;\;&0\;\;&1\;\;&0\;\;&0\;\;&0\;\;&0\;\;&0\;\;&0\;\;&0\;\;&0\;\;&1\;\;\\
0\;\;&0\;\;&0\;\;&0\;\;&0\;\;&1\;\;&0\;\;&0\;\;&0\;\;&0\;\;&-1\;\;&0\;\;&0\;\;&0\;\;\\
0\;\;&0\;\;&0\;\;&0\;\;&0\;\;&0\;\;&1\;\;&0\;\;&0\;\;&0\;\;&0\;\;&-1\;\;&0\;\;&0\;\;\\
0\;\;&0\;\;&0\;\;&0\;\;&0\;\;&0\;\;&0\;\;&1\;\;&0\;\;&0\;\;&0\;\;&0\;\;&-1\;\;&0\;\;\\
0\;\;&0\;\;&0\;\;&0\;\;&0\;\;&0\;\;&0\;\;&0\;\;&1\;\;&0\;\;&0\;\;&-1\;\;&-1\;\;&-1\;\;\\
0\;\;&0\;\;&0\;\;&0\;\;&0\;\;&0\;\;&0\;\;&0\;\;&0\;\;&1\;\;&0\;\;&-1\;\;&-1\;\;&-1\;\;\\
\end{array}\right)
\;,\nonumber\\
&&\vec{\vartheta_{_{5}}}^{\;T}=(\vartheta_{_{z_{_1}}},\cdots,\;\vartheta_{_{z_{_{14}}}})
\;,\nonumber\\
&&\mathbf{B_{_{5}}}^{\;T}=(-a_{_1},\;-a_{_2},\;-a_{_3},\;-a_{_4},\;-a_{_5},\;b_{_1}-1,\;b_{_2}-1
,\;b_{_3}-1,\;b_{_4}-1,\;b_{_5}-1)\;.
\label{GKZ18}
\end{eqnarray}
Here, the GKZ hypergeometric system for the three-loop vacuum integral  with five propagates is in keeping with that for the two-loop self energy with four propagators \cite{GKZ-2loop}.

Correspondingly the dual matrix $\mathbf{\tilde A_{_{5}}}$ of $\mathbf{A_{_{5}}}$ is
\begin{eqnarray}
&&\mathbf{\tilde A_{_{5}}}=\left(\begin{array}{cccccccccccccc}
-1\;\;&-1\;\;&0\;\;&0\;\;&0\;\;&1\;\;&0\;\;&0\;\;&0\;\;&0\;\;&1\;\;&0\;\;&0\;\;&0\;\;\\
-1\;\;&-1\;\;&-1\;\;&-1\;\;&0\;\;&0\;\;&1\;\;&0\;\;&1\;\;&1\;\;&0\;\;&1\;\;&0\;\;&0\;\;\\
-1\;\;&-1\;\;&-1\;\;&-1\;\;&0\;\;&0\;\;&0\;\;&1\;\;&1\;\;&1\;\;&0\;\;&0\;\;&1\;\;&0\;\;\\
-1\;\;&-1\;\;&0\;\;&0\;\;&-1\;\;&0\;\;&0\;\;&0\;\;&1\;\;&1\;\;&0\;\;&0\;\;&0\;\;&1\;\;
\end{array}\right).
\label{GKZ19}
\end{eqnarray}
The row vectors of the matrix $\mathbf{\tilde A_{_{5}}}$ induce the integer sublattice $\mathbf{B}$
which can be used to construct the formal solutions in hypergeometric series. Actually the integer sublattice $\mathbf{B}$
indicates that the solutions of the system should satisfy the equations in Eq. (\ref{GKZ17}) and the following
hyperbolic equations simultaneously
\begin{eqnarray}
&&{\partial^2\Phi_{_{5}}\over\partial z_{_1}\partial z_{_2}}={\partial^2\Phi_{_{5}}\over\partial z_{_6}\partial z_{_{11}}}
\;,\nonumber\\
&&{\partial^2\Phi_{_{5}}\over\partial z_{_7}\partial z_{_{12}}}={\partial^2\Phi_{_{5}}\over\partial z_{_8}\partial z_{_{13}}}
\;,\nonumber\\
&&{\partial^3\Phi_{_{5}}\over\partial z_{_3}\partial z_{_4}\partial z_{_{14}}}=
{\partial^3\Phi_{_{5}}\over\partial z_{_5}\partial z_{_{8}}\partial z_{_{13}}}
\;,\nonumber\\
&&{\partial^3\Phi_{_{5}}\over\partial z_{_5}\partial z_{_{6}}\partial z_{_{11}}}=
{\partial^3\Phi_{_{5}}\over\partial z_{_9}\partial z_{_{10}}\partial z_{_{14}}}
\;,\nonumber\\
&&{\partial^4\Phi_{_{5}}\over\partial z_{_3}\partial z_{_4}\partial z_{_6}\partial z_{_{11}}}=
{\partial^4\Phi_{_{5}}\over\partial z_{_8}\partial z_{_9}\partial z_{_{10}}\partial z_{_{13}}}\;.
\label{GKZ19-1}
\end{eqnarray}

Actually those partial differential equations is a Gr\"obner basis of the toric ideal of the matrix
$\mathbf{\tilde A_{_{5}}}$ presented Eq.~(\ref{GKZ19}). Defining the combined variables
\begin{eqnarray}
&&y_{_1}={z_{_6}z_{_{11}}\over z_{_1}z_{_2}}
\;,\quad y_{_2}={z_{_7}z_{_9}z_{_{10}}z_{_{12}}\over z_{_1}z_{_2}z_{_3}z_{_4}}
\;,\quad y_{_3}={z_{_8}z_{_9}z_{_{10}}z_{_{13}}\over z_{_1}z_{_2}z_{_3}z_{_4}}
\;,\quad y_{_4}={z_{_9}z_{_{10}}z_{_{14}}\over z_{_1}z_{_2}z_{_5}}\;,
\label{GKZ19-2}
\end{eqnarray}
we write the solutions satisfying Eq.~(\ref{GKZ17}) and Eq.~(\ref{GKZ19-1}) as
\begin{eqnarray}
&&\Phi_{_{5}}({\mathbf z})=\Big(\prod\limits_{i=1}^{14}z_{_i}^{\alpha_{_i}}\Big)
\varphi_{_{5}}(y_{_1},\;y_{_2},\;y_{_3},\;y_{_4})\;,
\label{GKZ19-3}
\end{eqnarray}
where $\vec{\alpha}^{\:T}=(\alpha_{_1},\;\alpha_{_2},\;\cdots,\;\alpha_{_{14}})$
denotes a sequence of complex number such that
\begin{eqnarray}
&&\mathbf{A_{_{5}}}\cdot\vec{\alpha}=\mathbf{B_{_{5}}}\;.
\label{GKZ19-4}
\end{eqnarray}
Substituting Eq.~(\ref{GKZ19-3}) in Eq.~(\ref{GKZ19-1}), we obtain five
independent partial differential equations $\hat{L}_{_i}\varphi_{_{5}}=0,\;\;(i=1,\cdots5)$,
where the linear partial differential operators $\hat{L}_{_i}$ can be seen in  Ref. \cite{GKZ-2loop}.

Through GKZ hypergeometric system in Eq.~(\ref{GKZ17}), total 536 hypergeometric functions are obtained in neighborhoods of origin and infinity, which can be seen in our work~\cite{GKZ-2loop}.
In certain nonempty intersections of corresponding convergent
regions of those hypergeometric series, the fundamental solution systems are composed by 30
linear independent hypergeometric functions. In other words, the analytical expression of the vacuum  integral can be formulated as a linear combination of those hypergeometric functions of corresponding fundamental solution system in certain convergent region.

In our previous works~\cite{Feng2020,GKZ-2loop}, we obtain GKZ hypergeometric systems of some one-loop and two-loop Feynman integrals, which show the algorithm and the obvious  hypergeometric series solutions for the one-loop and two-loop Feynman integrals. Meanwhile, we are researching GKZ hypergeometric systems of the three-loop vacuum Feynman integrals. Recently, the authors of the Refs.~\cite{Ananthanarayan2021,Ananthanarayan2022GKZ} give publicly available computer packages MBConicHulls~\cite{Ananthanarayan2021} and FeynGKZ~\cite{Ananthanarayan2022GKZ} to compute Feynman integrals in terms of hypergeometric functions, which are meaningful to improve computing efficiency. Through the package FeynGKZ~\cite{Ananthanarayan2022GKZ}, they give the examples of some one-loop and two-loop Feynman integrals, which are tested analytically by our previous work~\cite{Feng2020,GKZ-2loop}, as well as numerically using the package FIESTA~\cite{Smirnov-FIESTA5}.

Here, we also evaluate the three-loop vacuum Feynman integrals using FeynGKZ~\cite{Ananthanarayan2022GKZ}, which can be seen in the supplementary material. And then, through some transformations, the results obtained from the package FeynGKZ can be reduced to our results. For example, 8 terms of series solutions for the three-loop vacuum Feynman integral with four propagates are obtained from FeynGKZ, which are consistent with our simpler series solutions $\Phi_{_{[13\tilde{4}]}}^{(i)}(\alpha,z)\;(i=1,\cdots,8)$ in Appendix~\ref{app3}, through some transformations. In addition, our results also show the other 24 terms of series solutions $\Phi_{_{[134]}}^{(i)}(\alpha,z)$,  $\Phi_{_{[\tilde{1}34]}}^{(i)}(\alpha,z)$ and $\Phi_{_{[1\tilde{3}4]}}^{(i)}(\alpha,z)\;(i=1,\cdots,8)$ in the different convergent regions. The series solutions from GKZ hypergeometric systems are also tested numerically using FIESTA~\cite{Smirnov-FIESTA5}, which can be seen in the supplementary material.

In certain nonempty intersections of corresponding convergent regions of those hypergeometric series, the three-loop vacuum integrals can be formulated as a linear combination of those hypergeometric functions of corresponding fundamental solution system. In section \ref{sec-spe}, we also show that the combination coefficients how to be determined by the vacuum integral at some ordinary points or regular singularities, or the Mellin-Barnes representation of the vacuum Feynman integral.

\section{Special case for the three-loop vacuum integral with five propagates\label{sec-spe}}
\indent\indent

In order to elucidate how to obtain the analytical expression clearly, we assume the two nonzero virtual mass for the three-loop vacuum integral with five propagates.
The corresponding scalar integral of the special case for the three-loop vacuum diagram can be expressed as a linear combination of those
corresponding functionally independent Gauss functions or a linear combination of those
corresponding functionally independent Pochammer functions, respectively.

\subsection{The analytical expressions with $m_{_1}\neq0$, $m_{_5}\neq0$, $m_{_2}=m_{_3}=m_{_4}=0$\label{sec-spe-a}}
\indent\indent
Through above section, the GKZ hypergeometric system in this case can be simplified as
\begin{eqnarray}
&&\mathbf{A_{_{51}}}\cdot\vec{\vartheta}_{_{51}}\Phi_{_{51}}=\mathbf{B_{_{5}}}\Phi_{_{51}}\;,
\label{GKZ34a}
\end{eqnarray}
where the vector of Euler operators is defined as
\begin{eqnarray}
&&\vec{\vartheta}_{_{51}}^{\;T}=(\vartheta_{_{z_{_1}}},\;\vartheta_{_{z_{_{2}}}},\;\vartheta_{_{z_{_{3}}}},\;
\vartheta_{_{z_{_{4}}}},\;\vartheta_{_{z_{_{5}}}},\;\vartheta_{_{z_{_{6}}}},\;\vartheta_{_{z_{_{7}}}}
,\;\vartheta_{_{z_{_{8}}}},\;\vartheta_{_{z_{_{9}}}},\;\vartheta_{_{z_{_{10}}}},\;\vartheta_{_{z_{_{11}}}})\;,
\label{GKZ35a}
\end{eqnarray}
and the matrix $\mathbf{A}_{_{51}}$ is  obtained through deleting the 12th,
13th, and 14th columns of the matrix $\mathbf{A_{_{5}}}$:
\begin{eqnarray}
&&\mathbf{A}_{_{51}}=\left(\begin{array}{ccccccccccc}
1\;\;&0\;\;&0\;\;&0\;\;&0\;\;&0\;\;&0\;\;&0\;\;&0\;\;&0\;\;&1\;\\
0\;\;&1\;\;&0\;\;&0\;\;&0\;\;&0\;\;&0\;\;&0\;\;&0\;\;&0\;\;&1\;\\
0\;\;&0\;\;&1\;\;&0\;\;&0\;\;&0\;\;&0\;\;&0\;\;&0\;\;&0\;\;&0\;\\
0\;\;&0\;\;&0\;\;&1\;\;&0\;\;&0\;\;&0\;\;&0\;\;&0\;\;&0\;\;&0\;\\
0\;\;&0\;\;&0\;\;&0\;\;&1\;\;&0\;\;&0\;\;&0\;\;&0\;\;&0\;\;&0\;\\
0\;\;&0\;\;&0\;\;&0\;\;&0\;\;&1\;\;&0\;\;&0\;\;&0\;\;&0\;\;&-1\;\\
0\;\;&0\;\;&0\;\;&0\;\;&0\;\;&0\;\;&1\;\;&0\;\;&0\;\;&0\;\;&0\;\\
0\;\;&0\;\;&0\;\;&0\;\;&0\;\;&0\;\;&0\;\;&1\;\;&0\;\;&0\;\;&0\;\\
0\;\;&0\;\;&0\;\;&0\;\;&0\;\;&0\;\;&0\;\;&0\;\;&1\;\;&0\;\;&0\;\\
0\;\;&0\;\;&0\;\;&0\;\;&0\;\;&0\;\;&0\;\;&0\;\;&0\;\;&1\;\;&0\;\\
\end{array}\right)\;.
\label{GKZ36a}
\end{eqnarray}
Through Eq.~(\ref{GKZ34a}), one can have the relations
\begin{eqnarray}
&&\alpha_{_1}+\alpha_{_{11}}=-a_{_1}\;,
\quad \alpha_{_2}+\alpha_{_{11}}=-a_{_2}\;,
\quad \alpha_{_6}-\alpha_{_{11}}=b_{_1}-1\;,
\label{GKZ36a-1}
\end{eqnarray}
with $a_{_1}=4-D$, $a_{_2}=5-{3D\over2}$, $b_{_1}=2-{D\over2}$, and the other $\alpha_{_i}$ are zero.
The dual matrix $\mathbf{\tilde A}_{_{51}}$ of $\mathbf{A}_{_{51}}$ is
\begin{eqnarray}
&&\mathbf{\tilde A}_{_{51}}=\left(\begin{array}{ccccccccccc}
-1\;\;&-1\;\;&0\;\;&0\;\;&0\;\;&1\;\;&0\;\;&0\;\;&0\;\;&0\;\;&1\;\;
\end{array}\right)\;.
\label{GKZ37a}
\end{eqnarray}

The integer sublattice $\mathbf{B_{_{51}}}$ is determined by the dual matrix $\mathbf{\tilde A}_{_{51}}$ with $\mathbf{B_{_{51}}}=\mathbf{\tilde A}_{_{51}}$.
The integer sublattice $\mathbf{B_{_{51}}}$ induces the first hyperbolic equations
in Eq.~(\ref{GKZ19-1}), which implies that the system of fundamental solutions
is composed by two linear independent hypergeometric functions.
We take the set of column indices
$I_{_1}=[1,\cdots,10]$, which implies the exponent numbers $\alpha_{_{11}}=0$.
And then, the corresponding hypergeometric series solution can be written as
\begin{eqnarray}
&&\Phi_{_{[51]}}^{(1)}(y_{_1})=\;_{_2}F_{_1}
\left(\left.\begin{array}{cc}a_{_1},\;&a_{_2}\\
\;\;&b_{_1}\end{array}\right|y_{_1}\right)=\;_{_2}F_{_1}
\left(\left.\begin{array}{cc}4-D,\;&5-{3D\over2}\\
\;\;&2-{D\over2}\end{array}\right|y_{_1}\right)\;,
\label{GKZ38a1}
\end{eqnarray}
with $y_{_1}=x_{_1}=m_{_1}^2/m_{_5}^2$, and $_{_2}F_{_1}$ is Gauss function:
\begin{eqnarray}
&&_{_2}F_{_1}
\left(\left.\begin{array}{cc}a,\;&b\\
\;\;&c\end{array}\right|x\right)
=\;\sum\limits_{n=0}^\infty{(a)_n(b)_n
\over n!(c)_n}x^n\;,
\label{GKZ38a1-1}
\end{eqnarray}
with $(a)_n=\Gamma(a+n)/\Gamma(a)$. We also can take the set of column indices
$I_{_2}=[1,\cdots,5,7,\cdots,11]$, which implies the exponent numbers $\alpha_{_{6}}=0$.
And then, the another hypergeometric series solution can be written as
\begin{eqnarray}
&&\Phi_{_{[51]}}^{(2)}(y_{_1})=(y_{_1})^{1-b_{_1}}\;_{_2}F_{_1}
\left(\left.\begin{array}{cc}1+a_{_1}-b_{_1},\;&1+a_{_2}-b_{_1}\\
\;\;&2-b_{_1}\end{array}\right|y_{_1}\right)
\nonumber\\
&&\hspace{1.5cm}=
(y_{_1})^{D/2-1}\;_{_2}F_{_1}
\left(\left.\begin{array}{cc}3-{D\over2},\;&4-D\\
\;\;&{D\over2}\end{array}\right|y_{_1}\right)\;.
\label{GKZ38a2}
\end{eqnarray}
Here, the convergent region of the hypergeometric functions $\Phi_{_{[51]}}^{1,2}(y_{_1})$ is $|y_{_1}|<1$.
In the region $|y_{_1}|<1$, the scalar integral correspondingly is a linear combination of two fundamental solutions:
\begin{eqnarray}
&&\Phi_{_{51}}(y_{_1})=C_{_{[51]}}^{(1)}\Phi_{_{[51]}}^{(1)}(y_{_1})
+C_{_{[51]}}^{(2)}\Phi_{_{[51]}}^{(2)}(y_{_1})\;.
\label{GKZ38a3}
\end{eqnarray}

Multiplying one of the row vectors of the integer matrix $\mathbf{B}_{_{51}}$ by -1,
the induced integer matrix also can be chosen as a basis of the integer lattice
space of certain hypergeometric series.
And the corresponding system of
fundamental solutions for is similarly composed by two Gauss functions:
\begin{eqnarray}
&&\Phi_{_{[51]}}^{(3)}(y_{_1})=(y_{_1})^{-a_{_1}}\;_{_2}F_{_1}
\left(\left.\begin{array}{cc}a_{_1},\;&1+a_{_1}-b_{_1}\\
\;\;&1+a_{_1}-a_{_2}\end{array}\right|{1\over y_{_1}}\right)
\nonumber\\
&&\hspace{1.5cm}=
(y_{_1})^{D-4}\;_{_2}F_{_1}
\left(\left.\begin{array}{cc}4-D,\;&3-{D\over2}\\
\;\;&{D\over2}\end{array}\right|{1\over y_{_1}}\right)\;,
\nonumber\\
&&\Phi_{_{[51]}}^{(4)}(y_{_1})=(y_{_1})^{-a_{_2}}\;_{_2}F_{_1}
\left(\left.\begin{array}{cc}a_{_2},\;&1+a_{_2}-b_{_1}\\
\;\;&1-a_{_1}+a_{_2}\end{array}\right|{1\over y_{_1}}\right)
\nonumber\\
&&\hspace{1.5cm}=
(y_{_1})^{3D/2-5}\;_{_2}F_{_1}
\left(\left.\begin{array}{cc}5-{3D\over2},\;&4-D\\
\;\;&2-{D\over2}\end{array}\right|{1\over y_{_1}}\right)\;,
\label{GKZ39a}
\end{eqnarray}
which the convergent region is $|y_{_1}|>1$.
Correspondingly the scalar integral in the region $|y_{_1}|>1$ is a linear combination of two fundamental solutions:
\begin{eqnarray}
&&\Phi_{_{51}}(y_{_1})=C_{_{[51]}}^{(3)}\Phi_{_{[51]}}^{(3)}(y_{_1})
+C_{_{[51]}}^{(4)}\Phi_{_{[51]}}^{(4)}(y_{_1})\;.
\label{GKZ39a1}
\end{eqnarray}

As $m_{_1}^2\ll m_{_5}^2,\;m_{_2}=m_{_3}=m_{_4}=0$,
\begin{eqnarray}
&&I_{_1}=\Big(\Lambda_{_{\rm RE}}^2\Big)^{6-\frac{3D}{2}}\int{d^Dq_{_1}\over(2\pi)^D}
{d^Dq_{_2}\over(2\pi)^D}{d^Dq_{_3}\over(2\pi)^D}
{1\over(q_{_1}^2-m_{_1}^2)q_{_2}^2(q_{_1}+q_{_2}+q_{_3})^2
(q_{_1}+q_{_3})^2(q_{_3}^2-m_{_5}^2)}
\nonumber\\
&&\hspace{0.5cm}=I_{_{1,0}}+\cdots\;,
\label{GKZ40a}
\end{eqnarray}
where
\begin{eqnarray}
&&I_{_{1,0}}=\Big(\Lambda_{_{\rm RE}}^2\Big)^{6-\frac{3D}{2}}\int{d^Dq_{_1}\over(2\pi)^D}
{d^Dq_{_2}\over(2\pi)^D}{d^Dq_{_3}\over(2\pi)^D}
{1\over(q_{_1}^2)q_{_2}^2(q_{_1}+q_{_2}+q_{_3})^2
(q_{_1}+q_{_3})^2(q_{_3}^2-m_{_5}^2)}
\nonumber\\
&&\hspace{0.7cm}=
{im_{_5}^2\over(4\pi)^6}\Big({4\pi\Lambda_{_{\rm RE}}^2\over m_{_5}^2}\Big)^{6-{3D\over2}}
{\pi^2\Gamma(3-D)\over\sin{\pi D\over2}\sin{3\pi D\over2}}\;.
\label{GKZ41a}
\end{eqnarray}
This result indicates
\begin{eqnarray}
&&C_{_{[51]}}^{(1)}={im_{_5}^2\over(4\pi)^6}\Big({4\pi\Lambda_{_{\rm RE}}^2\over m_{_5}^2}\Big)^{6-{3D\over2}}
{\pi^2\Gamma(3-D)\over\sin{\pi D\over2}\sin{3\pi D\over2}}\;.
\label{GKZ41a-1}
\end{eqnarray}

As $m_{_1}^2\gg m_{_5}^2$ and $\;m_{_2}=m_{_3}=m_{_4}=0$, $I_{_1}=I_{_{1,\infty}}+\cdots\;$, where
\begin{eqnarray}
&&I_{_{1,\infty}}=\Big(\Lambda_{_{\rm RE}}^2\Big)^{6-\frac{3D}{2}}\int{d^Dq_{_1}\over(2\pi)^D}
{d^Dq_{_2}\over(2\pi)^D}{d^Dq_{_3}\over(2\pi)^D}
{1\over(q_{_1}^2-m_{_1}^2)q_{_2}^2(q_{_1}+q_{_2}+q_{_3})^2
(q_{_1}+q_{_3})^2q_{_3}^2}
\nonumber\\
&&\hspace{0.8cm}=
{im_{_1}^2\over(4\pi)^6}\Big({4\pi\Lambda_{_{\rm RE}}^2\over m_{_1}^2}\Big)^{6-{3D\over2}}
{\pi^2\Gamma(3-D)\over\sin{\pi D\over2}\sin{3\pi D\over2}}\;.
\label{GKZ44a}
\end{eqnarray}
This result indicates
\begin{eqnarray}
&&C_{_{[51]}}^{(4)}={im_{_5}^2\over(4\pi)^6}\Big({4\pi\Lambda_{_{\rm RE}}^2\over m_{_5}^2}\Big)^{6-{3D\over2}}
{\pi^2\Gamma(3-D)\over\sin{\pi D\over2}\sin{3\pi D\over2}}\;.
\label{GKZ44a-1}
\end{eqnarray}

Using the well known relation
\begin{eqnarray}
&&{\Gamma(a)\Gamma(b)\over\Gamma(c)}\;_{_2}F_{_1}
\left(\left.\begin{array}{cc}a,\;&b\\
\;\;&c\end{array}\right|z\right)
={\Gamma(a)\Gamma(b-a)\over\Gamma(c-a)}(-z)^{-a}\;_{_2}F_{_1}
\left(\left.\begin{array}{cc}a,\;&1+a-c\\
\;\;&1+a-b\end{array}\right|{1\over z}\right)
\nonumber\\
&&\hspace{5.0cm}
+{\Gamma(b)\Gamma(a-b)\over\Gamma(c-b)}(-z)^{-b}\;_{_2}F_{_1}
\left(\left.\begin{array}{cc}b,\;&1+b-c\\
\;\;&1-a+b\end{array}\right|{1\over z}\right)\;,
\label{GKZ44a-2}
\end{eqnarray}
we can obtain
\begin{eqnarray}
&&C_{_{[51]}}^{(3)}=C_{_{[51]}}^{(2)}={im_{_5}^2\over(4\pi)^6}\Big({4\pi\Lambda_{_{\rm RE}}^2\over m_{_5}^2}\Big)^{6-{3D\over2}}
{\pi^2\Gamma(3-D)\over\sin^2{\pi D\over2}}\;.
\label{GKZ44a-3}
\end{eqnarray}

Actually, the Mellin-Barnes representation
of the Feynman integral in this case can be obtained as
\begin{eqnarray}
&&\hspace{-0.5cm}U_{_5}=
{im_{_5}^2\over(4\pi)^6}\Big({4\pi\Lambda_{_{\rm RE}}^2\over m_{_5}^2}\Big)^{6-{3D\over2}}
{\pi\Gamma({D\over2}-1)\over(D-3)\sin\pi(2-{D\over2})\Gamma(3-{D\over2})}
\nonumber\\
&&\hspace{0.5cm}\times
{1\over2\pi i}\int_{-i\infty}^{+i\infty}ds_{_1}
\Big({m_{_1}^2\over m_{_5}^2}\Big)^{s_{_1}}{\Gamma(-s_{_1})
\Gamma({D\over2}-1-s_{_1})\Gamma(4-D+s_{_1})
\Gamma(5-{{3 D}\over2}+s_{_1})}\;.
\label{GKZ44a-4}
\end{eqnarray}
The residue of simple pole of $\Gamma(-s_{_1})$ provides $C_{_{[51]}}^{(1)}\Phi_{_{[51]}}^{(1)}(y_{_1})$,
that of simple pole of $\Gamma(D/2-1-s_{_1})$ provides $C_{_{[51]}}^{(2)}\Phi_{_{[51]}}^{(2)}(y_{_1})$,
that of simple pole of $\Gamma(4-D+s_{_1})$ provides $C_{_{[51]}}^{(3)}\Phi_{_{[51]}}^{(3)}(y_{_1})$,
and that of simple pole of $\Gamma(5-{{3 D}\over2}+s_{_1})$ provides $C_{_{[51]}}^{(4)}\Phi_{_{[51]}}^{(4)}(y_{_1})$,
respectively.

Combining the transformation
\begin{eqnarray}
&&\;_{_2}F_{_1}\left(\left.\begin{array}{cc}a,\;&b\\
\;\;&c\end{array}\right|z\right)
=(1-z)^{-b}\;_{_2}F_{_1}
\left(\left.\begin{array}{cc}c-a,\;&b\\
\;\;&c\end{array}\right|{z\over z-1}\right)
\label{GKZ44a-5}
\end{eqnarray}
with the relation Eq.~(\ref{GKZ44a-2}), one can derive the analytic expression
in the neighborhood of $y_{_1}=1$, i.e. the analytic expression of
the neighborhood of the threshold. Note that the threshold exactly coincides
with the regular singularity $z=1$ of the PDEs satisfied by the Feynman integral
in this special case.
Taking Feynman integral as a function of some subvarieties of Grassmannians~\cite{Grassmannians},
we can get the similar relations of Eq. (\ref{GKZ44a-5}) among the generalized hypergeometric functions
through the algorithm in combinatorial geometry.

\subsection{The analytical expressions with $m_{_4}\neq0$,  $m_{_5}\neq0$, $m_{_1}=m_{_2}=m_{_3}=0$\label{sec-spe-b}}
\indent\indent
In this case, the GKZ hypergeometric system is simplified as
\begin{eqnarray}
&&\mathbf{A_{_{54}}}\cdot\vec{\vartheta}_{_{54}}\Phi_{_{54}}=\mathbf{B_{_{5}}}\Phi_{_{54}}\;,
\label{GKZ34b}
\end{eqnarray}
where the vector of Euler operators is defined as
\begin{eqnarray}
&&\vec{\vartheta}_{_{54}}^{\;T}=(\vartheta_{_{z_{_1}}},\;\vartheta_{_{z_{_{2}}}},\;\vartheta_{_{z_{_{3}}}},\;
\vartheta_{_{z_{_{4}}}},\;\vartheta_{_{z_{_{5}}}},\;\vartheta_{_{z_{_{6}}}},\;\vartheta_{_{z_{_{7}}}}
,\;\vartheta_{_{z_{_{8}}}},\;\vartheta_{_{z_{_{9}}}},\;\vartheta_{_{z_{_{10}}}},\;\vartheta_{_{z_{_{14}}}})\;,
\label{GKZ35b}
\end{eqnarray}
and the matrix $\mathbf{A}_{_{54}}$ is
\begin{eqnarray}
&&\mathbf{A}_{_{54}}=\left(\begin{array}{ccccccccccc}
1\;\;&0\;\;&0\;\;&0\;\;&0\;\;&0\;\;&0\;\;&0\;\;&0\;\;&0\;\;&1\;\;\\
0\;\;&1\;\;&0\;\;&0\;\;&0\;\;&0\;\;&0\;\;&0\;\;&0\;\;&0\;\;&1\;\;\\
0\;\;&0\;\;&1\;\;&0\;\;&0\;\;&0\;\;&0\;\;&0\;\;&0\;\;&0\;\;&0\;\;\\
0\;\;&0\;\;&0\;\;&1\;\;&0\;\;&0\;\;&0\;\;&0\;\;&0\;\;&0\;\;&0\;\;\\
0\;\;&0\;\;&0\;\;&0\;\;&1\;\;&0\;\;&0\;\;&0\;\;&0\;\;&0\;\;&1\;\;\\
0\;\;&0\;\;&0\;\;&0\;\;&0\;\;&1\;\;&0\;\;&0\;\;&0\;\;&0\;\;&0\;\;\\
0\;\;&0\;\;&0\;\;&0\;\;&0\;\;&0\;\;&1\;\;&0\;\;&0\;\;&0\;\;&0\;\;\\
0\;\;&0\;\;&0\;\;&0\;\;&0\;\;&0\;\;&0\;\;&1\;\;&0\;\;&0\;\;&0\;\;\\
0\;\;&0\;\;&0\;\;&0\;\;&0\;\;&0\;\;&0\;\;&0\;\;&1\;\;&0\;\;&-1\;\;\\
0\;\;&0\;\;&0\;\;&0\;\;&0\;\;&0\;\;&0\;\;&0\;\;&0\;\;&1\;\;&-1\;\;\\
\end{array}\right)\;.
\label{GKZ36b}
\end{eqnarray}

The integer sublattice is determined by
the dual matrix $\mathbf{\tilde A}_{_{54}}$ of $\mathbf{A}_{_{54}}$,
\begin{eqnarray}
&&\mathbf{\tilde A}_{_{54}}=\left(\begin{array}{ccccccccccc}
-1\;\;&-1\;\;&0\;\;&0\;\;&-1\;\;&0\;\;&0\;\;&0\;\;&1\;\;&1\;\;&1\;\;
\end{array}\right).
\label{GKZ37b}
\end{eqnarray}
In the region $|y_{_4}|<1$, the corresponding system of
fundamental solutions is composed by three Pochammer functions $_{_3}F_{_2}$,
which are simplified as Gauss functions under the special exponents
of Eq. (\ref{GKZ6-1}):
\begin{eqnarray}
&&\Phi_{_{[54]}}^{(1)}(y_{_4})=\;_{_3}F_{_2}
\left(\left.\begin{array}{ccc}a_{_1},\;&a_{_2},\;&a_{_5}\\
\;\;&b_{_4},\;&b_{_5}\end{array}\right|y_{_4}\right)
=\;_{_2}F_{_1}\left(\left.\begin{array}{cc}1,\;&5-{3D\over2}\\
\;\;&3-{D\over2}\end{array}\right|y_{_4}\right)\;,
\nonumber\\
&&\Phi_{_{[54]}}^{(2)}(y_{_4})=(y_{_4})^{1-b_{_5}}\;_{_3}F_{_2}
\left(\left.\begin{array}{ccc}1+a_{_1}-b_{_5},\;&1+a_{_2}-b_{_5},\;&1+a_{_5}-b_{_5}\\
\;\;&2-b_{_5},\;&1+b_{_4}-b_{_5}\end{array}\right|y_{_4}\right)
\nonumber\\
&&\hspace{1.5cm}=
(y_{_4})^{D-3}\;_{_2}F_{_1}\left(\left.\begin{array}{cc}1,\;&2-{D\over2}\\
\;\;&{D\over2}\end{array}\right|y_{_4}\right)\;,
\nonumber\\
&&\Phi_{_{[54]}}^{(3)}(y_{_4})=(y_{_4})^{1-b_{_4}}\;_{_3}F_{_2}
\left(\left.\begin{array}{ccc}1+a_{_1}-b_{_4},\;&1+a_{_2}-b_{_4},\;&1+a_{_5}-b_{_4}\\
\;\;&2-b_{_4},\;&1-b_{_4}+b_{_5}\end{array}\right|y_{_4}\right)
\nonumber\\
&&\hspace{1.5cm}=
(y_{_4})^{D/2-2}(1-y_{_4})^{D-3}\;,
\label{GKZ38b}
\end{eqnarray}
with $y_{_4}=x_{_4}=m_{_4}^2/m_{_5}^2$, and $_{_3}F_{_2}$ is three Pochammer function:
\begin{eqnarray}
&&_{_3}F_{_2}
\left(\left.\begin{array}{ccc}a,\;&b,\;&c\\
\;\;&d,\;&e\end{array}\right|x\right)
=\;\sum\limits_{n=0}^\infty{(a)_n(b)_n(c)_n
\over n!(d)_n(e)_n}x^n\;.
\label{GKZ38b-1-1}
\end{eqnarray}
Correspondingly the Feynman integral is formulated as a linear combination
\begin{eqnarray}
&&\Phi_{_{54}}(y_{_4})=C_{_{[54]}}^{(1)}\Phi_{_{[54]}}^{(1)}(y_{_4})
+C_{_{[54]}}^{(2)}\Phi_{_{[54]}}^{(2)}(y_{_4})+C_{_{[54]}}^{(3)}\Phi_{_{[54]}}^{(3)}(y_{_4}),
\label{GKZ38b-1}
\end{eqnarray}
in the region $|y_{_4}|<1$.

In the region $|y_{_4}|>1$, the corresponding system of
fundamental solutions is similarly composed by three Pochammer functions $_{_3}F_{_2}$,
which are simplified as Gauss functions under the special exponents
of Eq. (\ref{GKZ6-1}):
\begin{eqnarray}
&&\Phi_{_{[54]}}^{(4)}(y_{_4})=(y_{_4})^{-a_{_1}}\;_{_3}F_{_2}
\left(\left.\begin{array}{ccc}a_{_1},\;&1+a_{_1}-b_{_4},\;&1+a_{_1}-b_{_5}\\
\;\;&1+a_{_1}-a_{_2},\;&1+a_{_1}-a_{_5}\end{array}\right|{1\over y_{_4}}\right)
\nonumber\\
&&\hspace{1.5cm}=
(y_{_4})^{D-4}\;_{_2}F_{_1}
\left(\left.\begin{array}{cc}1,\;&2-{D\over2}\\
\;\;&{D\over2}\end{array}\right|{1\over y_{_4}}\right)\;,
\nonumber\\
&&\Phi_{_{[54]}}^{(5)}(y_{_4})=(y_{_4})^{-a_{_5}}\;_{_3}F_{_2}
\left(\left.\begin{array}{ccc}a_{_5},\;&1+a_{_5}-b_{_4},\;&1+a_{_5}-b_{_5}\\
\;\;&1-a_{_2}+a_{_5},\;&1-a_{_1}+a_{_5}\end{array}\right|{1\over y_{_4}}\right)
\nonumber\\
&&\hspace{1.5cm}=
{1\over y_{_4}}\;_{_2}F_{_1}
\left(\left.\begin{array}{cc}1,\;&{D\over2}-1\\
\;\;&{3D\over2}-3\end{array}\right|{1\over y_{_4}}\right)\;,
\nonumber\\
&&\Phi_{_{[54]}}^{(6)}(y_{_4})=(y_{_4})^{-a_{_2}}\;_{_3}F_{_2}
\left(\left.\begin{array}{ccc}a_{_2},\;&1+a_{_2}-b_{_4},\;&1+a_{_2}-b_{_5}\\
\;\;&1-a_{_1}+a_{_2},\;&1+a_{_2}-a_{_5}\end{array}\right|{1\over y_{_4}}\right)
\nonumber\\
&&\hspace{1.5cm}=
(y_{_4})^{D/2-2}(1-y_{_4})^{D-3}\;.
\label{GKZ39b}
\end{eqnarray}
Correspondingly the Feynman integral is formulated as a linear combination
\begin{eqnarray}
&&\Phi_{_{54}}(y_{_4})=C_{_{[54]}}^{(4)}\Phi_{_{[54]}}^{(4)}(y_{_4})
+C_{_{[54]}}^{(5)}\Phi_{_{[54]}}^{(5)}(y_{_4})+C_{_{[54]}}^{(6)}\Phi_{_{[54]}}^{(6)}(y_{_4}),
\label{GKZ39b-1}
\end{eqnarray}
in the region $|y_{_4}|>1$.

As $m_{_4}^2\gg  m_{_5}^2,\;m_{_1}=m_{_2}=m_{_3}=0$,
\begin{eqnarray}
&&I_{_4}=\Big(\Lambda_{_{\rm RE}}^2\Big)^{6-\frac{3D}{2}}\int{d^Dq_{_1}\over(2\pi)^D}
{d^Dq_{_2}\over(2\pi)^D}{d^Dq_{_3}\over(2\pi)^D}
{1\over q_{_1}^2q_{_2}^2(q_{_1}+q_{_2}+q_{_3})^2
((q_{_1}+q_{_3})^2-m_{_4}^2)(q_{_3}^2-m_{_5}^2)}
\nonumber\\
&&\hspace{0.5cm}=I_{_{4,\infty}}+\cdots\;,
\label{GKZ42b}
\end{eqnarray}
where
\begin{eqnarray}
&&I_{_{4,\infty}}=\Big(\Lambda_{_{\rm RE}}^2\Big)^{6-\frac{3D}{2}}\int{d^Dq_{_1}\over(2\pi)^D}
{d^Dq_{_2}\over(2\pi)^D}{d^Dq_{_3}\over(2\pi)^D}
{1\over q_{_1}^2q_{_2}^2(q_{_1}+q_{_2}+q_{_3})^2
((q_{_1}+q_{_3})^2-m_{_4}^2)q_{_3}^2}
\nonumber\\
&&\hspace{0.8cm}={im_{_4}^2\over(4\pi)^6}\Big({4\pi\Lambda_{_{\rm RE}}^2\over m_{_4}^2}\Big)^{6-{3D\over2}}
{\pi^3\over\sin^2{\pi D\over2}\sin{3\pi D\over2}}\;.
\label{GKZ44b}
\end{eqnarray}

The results of Eq.~(\ref{GKZ41a}) and Eq.~(\ref{GKZ44b}) induce
\begin{eqnarray}
&&C_{_{[54]}}^{(1)}={im_{_5}^2\over(4\pi)^6}\Big({4\pi\Lambda_{_{\rm RE}}^2\over m_{_5}^2}\Big)^{6-{3D\over2}}
{\pi^2\Gamma(3-D)\over\sin{\pi D\over2}\sin{3\pi D\over2}}\;,
\nonumber\\
&&C_{_{[54]}}^{(3)}=C_{_{[54]}}^{(6)}={im_{_5}^2\over(4\pi)^6}\Big({4\pi\Lambda_{_{\rm RE}}^2\over m_{_5}^2}\Big)^{6-{3D\over2}}
{\pi^3\over\sin^2{\pi D\over2}\sin{3\pi D\over2}}\;.
\label{GKZ44b-1}
\end{eqnarray}
Furthermore, the relation Eq.~(\ref{GKZ44a-2}) indicates
\begin{eqnarray}
&&C_{_{[54]}}^{(5)}={im_{_5}^2\over(4\pi)^6}\Big({4\pi\Lambda_{_{\rm RE}}^2\over m_{_5}^2}\Big)^{6-{3D\over2}}
{\pi\Gamma(3-D)\over \sin{3\pi D\over2}\Gamma({3D\over2}-3)}\;.
\label{GKZ44b-2}
\end{eqnarray}

The combination coefficients also can be obtained through the Mellin-Barnes
representation of the Feynman integral of this special case:
\begin{eqnarray}
&&\hspace{-0.5cm}U_{_5}=
{im_{_5}^2\over(4\pi)^6}\Big({4\pi\Lambda_{_{\rm RE}}^2\over m_{_5}^2}\Big)^{6-{3D\over2}}
{\pi\Gamma^2({D\over2}-1)\over\sin\pi(2-{D\over2})\Gamma(D-2)}
{1\over2\pi i}\int_{-i\infty}^{+i\infty}ds_{_4}\Big({m_{_4}^2\over m_{_5}^2}\Big)^{s_{_4}}
\nonumber\\
&&\hspace{0.5cm}\times
{\Gamma(-s_{_4})\Gamma(1+s_{_4})
\Gamma(D-3-s_{_4})\Gamma(4-D+s_{_4})
\Gamma(5-{{3 D}\over2}+s_{_4})
\over \Gamma(3-{D\over2}+s_{_4})}\;.
\label{GKZ44b-3}
\end{eqnarray}
The residue of simple pole of $\Gamma(-s_{_4})$ provides $C_{_{[54]}}^{(1)}\Phi_{_{[54]}}^{(1)}(y_{_4})$, that of simple pole of $\Gamma(1+s_{_4})$ provides $C_{_{[54]}}^{(5)}\Phi_{_{[54]}}^{(5)}(y_{_4})$,
that of simple pole of $\Gamma(D-3-s_{_4})$ provides $C_{_{[54]}}^{(2)}\Phi_{_{[54]}}^{(2)}(y_{_4})$,
that of simple pole of $\Gamma(4-D+s_{_4})$ provides $C_{_{[54]}}^{(4)}\Phi_{_{[54]}}^{(4)}(y_{_4})$,
and that of simple pole of $\Gamma(5-{{3 D}\over2}+s_{_4})$ provides $C_{_{[54]}}^{(3)}\Phi_{_{[54]}}^{(3)}(y_{_4})$ and  $C_{_{[54]}}^{(6)}\Phi_{_{[54]}}^{(6)}(y_{_4})$,
respectively.
And the residue of the simple pole of $\Gamma(D-3-s_{_4})$ and $\Gamma(4-D+s_{_4})$ can induce
\begin{eqnarray}
&&C_{_{[54]}}^{(2)}=C_{_{[54]}}^{(4)}={im_{_5}^2\over(4\pi)^6}\Big({4\pi\Lambda_{_{\rm RE}}^2\over m_{_5}^2}\Big)^{6-{3D\over2}}
{\pi^2\Gamma(3-D)\over\sin^2{\pi D\over2}}\;.
\label{GKZ44b-4}
\end{eqnarray}

\subsection{The analytical expressions with $m_{_2}\neq0$, $m_{_5}\neq0$, $m_{_1}=m_{_3}=m_{_4}=0$\label{sec-spe-c}}
\indent\indent
In the parameter space, the GKZ hypergeometric system is simplified as
\begin{eqnarray}
&&\mathbf{A}_{_{52}}\cdot\vec{\vartheta}_{_{52}}\Phi_{_{52}}=\mathbf{B_{_{5}}}\Phi_{_{52}}\;,
\label{GKZ34c}
\end{eqnarray}
where the vector of Euler operators is
\begin{eqnarray}
&&\vec{\vartheta}_{_{52}}^{\;T}=(\vartheta_{_{z_{_1}}},\;\vartheta_{_{z_{_{2}}}},\;\vartheta_{_{z_{_{3}}}},\;
\vartheta_{_{z_{_{4}}}},\;\vartheta_{_{z_{_{5}}}},\;\vartheta_{_{z_{_{6}}}},\;\vartheta_{_{z_{_{7}}}}
,\;\vartheta_{_{z_{_{8}}}},\;\vartheta_{_{z_{_{9}}}},\;\vartheta_{_{z_{_{10}}}},\;\vartheta_{_{z_{_{12}}}})\;,
\label{GKZ35c}
\end{eqnarray}
and the matrix $\mathbf{A}_{_{52}}$ is
\begin{eqnarray}
&&\mathbf{A}_{_{52}}=\left(\begin{array}{ccccccccccc}
1\;\;&0\;\;&0\;\;&0\;\;&0\;\;&0\;\;&0\;\;&0\;\;&0\;\;&0\;\;&1\;\;\\
0\;\;&1\;\;&0\;\;&0\;\;&0\;\;&0\;\;&0\;\;&0\;\;&0\;\;&0\;\;&1\;\;\\
0\;\;&0\;\;&1\;\;&0\;\;&0\;\;&0\;\;&0\;\;&0\;\;&0\;\;&0\;\;&1\;\;\\
0\;\;&0\;\;&0\;\;&1\;\;&0\;\;&0\;\;&0\;\;&0\;\;&0\;\;&0\;\;&1\;\;\\
0\;\;&0\;\;&0\;\;&0\;\;&1\;\;&0\;\;&0\;\;&0\;\;&0\;\;&0\;\;&0\;\;\\
0\;\;&0\;\;&0\;\;&0\;\;&0\;\;&1\;\;&0\;\;&0\;\;&0\;\;&0\;\;&0\;\;\\
0\;\;&0\;\;&0\;\;&0\;\;&0\;\;&0\;\;&1\;\;&0\;\;&0\;\;&0\;\;&-1\;\;\\
0\;\;&0\;\;&0\;\;&0\;\;&0\;\;&0\;\;&0\;\;&1\;\;&0\;\;&0\;\;&0\;\;\\
0\;\;&0\;\;&0\;\;&0\;\;&0\;\;&0\;\;&0\;\;&0\;\;&1\;\;&0\;\;&-1\;\;\\
0\;\;&0\;\;&0\;\;&0\;\;&0\;\;&0\;\;&0\;\;&0\;\;&0\;\;&1\;\;&-1\;\;\\
\end{array}\right)\;.
\label{GKZ36c}
\end{eqnarray}

The integer sublattice is determined by
the dual matrix $\mathbf{\tilde A}_{_{52}}$ of $\mathbf{A}_{_{52}}$,
\begin{eqnarray}
&&\mathbf{\tilde A}_{_{52}}=\left(\begin{array}{ccccccccccc}
-1\;\;&-1\;\;&-1\;\;&-1\;\;&0\;\;&0\;\;&1\;\;&0\;\;&1\;\;&1\;\;&1\;\;
\end{array}\right).
\label{GKZ37c}
\end{eqnarray}
In the region $|y_{_2}|<1$, the corresponding system of
fundamental solutions is composed by four Pochammer functions $_{_4}F_{_3}$,
which are simplified as Gauss functions or three Pochammer functions under the special exponents
of Eq. (\ref{GKZ6-1}):
\begin{eqnarray}
&&\Phi_{_{[52]}}^{(1)}(y_{_2})=\;_{_4}F_{_3}
\left(\left.\begin{array}{cccc}a_{_1},\;&a_{_2},\;&a_{_3},\;&a_{_4}\\
\;\;&b_{_2},\;&b_{_4},\;&b_{_5}\end{array}\right|y_{_2}\right)
=\;_{_2}F_{_1}\left(\left.\begin{array}{cc}3-D,\;&5-{3D\over2}\\
\;\;&3-{D\over2}\end{array}\right|y_{_2}\right)\;,
\nonumber\\
&&\Phi_{_{[52]}}^{(2)}(y_{_2})=(-y_{_2})^{1-b_{_2}}\;_{_4}F_{_3}
\left(\left.\begin{array}{cccc}1+a_{_1}-b_{_2},\;&1+a_{_2}-b_{_2},\;&1+a_{_3}-b_{_2},\;&1+a_{_4}-b_{_2}\\
\;\;&2-b_{_2},\;&1-b_{_2}+b_{_4},\;&1-b_{_2}+b_{_5}\end{array}\right|y_{_2}\right)
\nonumber\\
&&\hspace{1.5cm}=
(y_{_2})^{D/2-1}\;_{_3}F_{_2}
\left(\left.\begin{array}{ccc}1,\;&2-{D\over2},\;&4-D\\
\;\;&2,\;&{D\over2}\end{array}\right|y_{_2}\right)
\;,\nonumber\\
&&\Phi_{_{[52]}}^{(3)}(y_{_2})=(y_{_2})^{1-b_{_4}}\;_{_4}F_{_3}
\left(\left.\begin{array}{cccc}1+a_{_1}-b_{_4},\;&1+a_{_2}-b_{_4},\;&1+a_{_3}-b_{_4},\;&1+a_{_4}-b_{_4}\\
\;\;&2-b_{_4},\;&1+b_{_2}-b_{_4},\;&1-b_{_4}+b_{_5}\end{array}\right|y_{_2}\right)
\nonumber\\
&&\hspace{1.5cm}=
(y_{_2})^{D/2-2}\;_{_2}F_{_1}\left(\left.\begin{array}{cc}1-{D\over2},\;&3-D\\
\;\;&{D\over2}-1\end{array}\right|y_{_2}\right)
\;,\nonumber\\
&&\Phi_{_{[52]}}^{(4)}(y_{_2})=(y_{_2})^{1-b_{_5}}\;_{_4}F_{_3}
\left(\left.\begin{array}{cccc}1+a_{_1}-b_{_5},\;&1+a_{_2}-b_{_5},\;&1+a_{_3}-b_{_5},\;&1+a_{_4}-b_{_5}\\
\;\;&2-b_{_5},\;&1+b_{_2}-b_{_5},\;&1+b_{_4}-b_{_5}\end{array}\right|y_{_2}\right)
\nonumber\\
&&\hspace{1.5cm}=
(y_{_2})^{D-3}\;,
\label{GKZ38c}
\end{eqnarray}
where $y_{_2}=x_{_2}=m_{_2}^2/m_{_5}^2$, and $_{_4}F_{_3}$ is four Pochammer function:
\begin{eqnarray}
&&_{_4}F_{_3}
\left(\left.\begin{array}{cccc}a,\;&b,\;&c,\;&d\\
\;\;&e,\;&f,\;&g\end{array}\right|x\right)
=\;\sum\limits_{n=0}^\infty{(a)_n(b)_n(c)_n(d)_n
\over n!(e)_n(f)_n(g)_n}x^n\;.
\label{GKZ38c-1-1}
\end{eqnarray}
Correspondingly the Feynman integral is formulated as a linear combination
\begin{eqnarray}
&&\Phi_{_{[52]}}(y_{_2})=C_{_{[52]}}^{(1)}\Phi_{_{[52]}}^{(1)}(y_{_2})
+C_{_{[52]}}^{(2)}\Phi_{_{[52]}}^{(2)}(y_{_2})
+C_{_{[52]}}^{(3)}\Phi_{_{[52]}}^{(3)}(y_{_2})
+C_{_{[52]}}^{(4)}\Phi_{_{[52]}}^{(4)}(y_{_2}),
\label{GKZ38c-1}
\end{eqnarray}
in the region $|y_{_2}|<1$.

In the region $|y_{_2}|>1$, the corresponding system of
fundamental solutions is similarly composed by four Pochammer functions $_{_4}F_{_3}$,
which are simplified under the special exponents
of Eq. (\ref{GKZ6-1}):
\begin{eqnarray}
&&\Phi_{_{[52]}}^{(5)}(y_{_2})=(y_{_2})^{-a_{_1}}\;_{_4}F_{_3}
\left(\left.\begin{array}{cccc}a_{_1},\;&1+a_{_1}-b_{_2},\;&1+a_{_1}-b_{_4},\;&1+a_{_1}-b_{_5}\\
\;\;&1+a_{_1}-a_{_2},\;&1+a_{_1}-a_{_3},\;&1+a_{_1}-a_{_4}\end{array}\right|{1\over y_{_2}}\right)
\nonumber\\
&&\hspace{1.5cm}=
(y_{_2})^{D-4}\;_{_3}F_{_2}
\left(\left.\begin{array}{ccc}1,\;&4-D,\;&2-{D\over2}\\
\;\;&2,\;&{D\over2}\end{array}\right|{1\over y_{_2}}\right)
\;,\nonumber\\
&&\Phi_{_{[52]}}^{(6)}(y_{_2})=(y_{_2})^{-a_{_3}}\;_{_4}F_{_3}
\left(\left.\begin{array}{cccc}a_{_3},\;&1+a_{_3}-b_{_2},\;&1+a_{_3}-b_{_4},\;&1+a_{_3}-b_{_5}\\
\;\;&1-a_{_2}+a_{_3},\;&1-a_{_1}+a_{_3},\;&1+a_{_3}-a_{_4}\end{array}\right|{1\over y_{_2}}\right)
\nonumber\\
&&\hspace{1.5cm}=
(y_{_2})^{D/2-2}
\;,\nonumber\\
&&\Phi_{_{[52]}}^{(7)}(y_{_2})=(y_{_2})^{-a_{_2}}\;_{_4}F_{_3}
\left(\left.\begin{array}{cccc}a_{_2},\;&1+a_{_2}-b_{_2},\;&1+a_{_2}-b_{_4},\;&1+a_{_2}-b_{_5}\\
\;\;&1-a_{_1}+a_{_2},\;&1+a_{_2}-a_{_3},\;&1+a_{_2}-a_{_4}\end{array}\right|{1\over y_{_2}}\right)
\nonumber\\
&&\hspace{1.5cm}=
(y_{_2})^{3D/2-5}\;_{_2}F_{_1}
\left(\left.\begin{array}{cc}3-D,\;&5-{3D\over2}\\
\;\;&3-{D\over2}\end{array}\right|{1\over y_{_2}}\right)
\;,\nonumber\\
&&\Phi_{_{[52]}}^{(8)}(y_{_2})=(y_{_2})^{-a_{_4}}\;_{_4}F_{_3}
\left(\left.\begin{array}{cccc}a_{_4},\;&1+a_{_4}-b_{_2},\;&1+a_{_4}-b_{_4},\;&1+a_{_4}-b_{_5}\\
\;\;&1-a_{_2}+a_{_4},\;&1-a_{_1}+a_{_4},\;&1-a_{_3}+a_{_4}\end{array}\right|{1\over y_{_2}}\right)
\nonumber\\
&&\hspace{1.5cm}=
(y_{_2})^{D-3}\;_{_2}F_{_1}
\left(\left.\begin{array}{cc}3-D,\;&1-{D\over2}\\
\;\;&{D\over2}-1\end{array}\right|{1\over y_{_2}}\right)\;.
\label{GKZ39c}
\end{eqnarray}
Similarly the Feynman integral is formulated as a linear combination
\begin{eqnarray}
&&\Phi_{_{[52]}}(y_{_2})=C_{_{[52]}}^{(5)}\Phi_{_{[52]}}^{(5)}(y_{_2})
+C_{_{[52]}}^{(6)}\Phi_{_{[52]}}^{(6)}(y_{_2})+C_{_{[52]}}^{(7)}\Phi_{_{[52]}}^{(7)}(y_{_2})
+C_{_{[52]}}^{(8)}\Phi_{_{[52]}}^{(8)}(y_{_2}),
\label{GKZ39c-1}
\end{eqnarray}
in the region $|y_{_2}|>1$.

As $m_{_2}^2\gg  m_{_5}^2,\;m_{_1}=m_{_3}=m_{_4}=0$,
\begin{eqnarray}
&&I_{_2}=\Big(\Lambda_{_{\rm RE}}^2\Big)^{6-\frac{3D}{2}}\int{d^Dq_{_1}\over(2\pi)^D}
{d^Dq_{_2}\over(2\pi)^D}{d^Dq_{_3}\over(2\pi)^D}
{1\over q_{_1}^2(q_{_2}^2-m_{_4}^2)(q_{_1}+q_{_2}+q_{_3})^2
(q_{_1}+q_{_3})^2(q_{_3}^2-m_{_5}^2)}
\nonumber\\
&&\hspace{0.5cm}=I_{_{2,\infty}}+\cdots\;,
\label{GKZ39c-1-1}
\end{eqnarray}
where
\begin{eqnarray}
&&I_{_{2,\infty}}=\Big(\Lambda_{_{\rm RE}}^2\Big)^{6-\frac{3D}{2}}\int{d^Dq_{_1}\over(2\pi)^D}
{d^Dq_{_2}\over(2\pi)^D}{d^Dq_{_3}\over(2\pi)^D}
{1\over q_{_1}^2(q_{_2}^2-m_{_4}^2)(q_{_1}+q_{_2}+q_{_3})^2
(q_{_1}+q_{_3})^2q_{_3}^2}
\nonumber\\
&&\hspace{0.8cm}={im_{_2}^2\over(4\pi)^6}\Big({4\pi\Lambda_{_{\rm RE}}^2\over m_{_2}^2}\Big)^{6-{3D\over2}}
{\pi^2\Gamma(3-D)\over\sin{\pi D\over2}\sin{3\pi D\over2}}\;.
\label{GKZ39c-1-2}
\end{eqnarray}
The regular value of the Feynman integral
as $m_{_5}^2\neq0$, $m_{_i}^2=0,\;(i=1,\cdots,4)$ in Eq.~(\ref{GKZ41a})
and that of  the Feynman integral
as $m_{_2}^2\neq0$, $m_{_i}^2=0,\;(i=1,3,4,5)$ in Eq.~(\ref{GKZ39c-1-2}) induce
\begin{eqnarray}
&&C_{_{[52]}}^{(1)}=C_{_{[52]}}^{(7)}={im_{_5}^2\over(4\pi)^6}\Big({4\pi\Lambda_{_{\rm RE}}^2\over m_{_5}^2}\Big)^{6-{3D\over2}}
{\pi^2\Gamma(3-D)\over\sin{\pi D\over2}\sin{3\pi D\over2}}\;.
\label{GKZ39c-3}
\end{eqnarray}
Because there is no relation among the Pochammer functions $\;_{_3}F_{_2}$
which is similar to the relation among the Gauss functions $\;_{_2}F_{_1}$
presented in Eq.~(\ref{GKZ44a-2}), one cannot derive the constraints on the
combination coefficients similar to Eq.~(\ref{GKZ44a-3}).
However, the combination coefficients can be obtained through the Mellin-Barnes
representation of the Feynman integral of this special case:
\begin{eqnarray}
&&\hspace{-0.5cm}U_{_5}=
{im_{_5}^2\over(4\pi)^6}\Big({4\pi\Lambda_{_{\rm RE}}^2\over m_{_5}^2}\Big)^{6-{3D\over2}}
{\Gamma^2({D\over2}-1)\over2\pi i}\int_{-i\infty}^{+i\infty}ds_{_2}\Big({m_{_2}^2\over m_{_5}^2}\Big)^{s_{_2}}\Gamma(-s_{_2})
\nonumber\\
&&\hspace{0.5cm}\times
{\Gamma({D\over2}-1-s_{_2})\Gamma(2-{ D\over2}+s_{_2})
\Gamma(D-3-s_{_2})\Gamma(4-D+s_{_2})
\Gamma(5-{{3 D}\over2}+s_{_2})
\over \Gamma(D-2-s_{_2})\Gamma(3-{D\over2}+s_{_2})}\;,
\label{GKZ39c-4}
\end{eqnarray}
where the coefficient $C_{_{[52]}}^{(1)}$ is determined from the residue of the simple
pole of $\Gamma(-s_{_2})$, and the coefficient $C_{_{[52]}}^{(7)}$ is determined from
the residue of the simple pole of $\Gamma(5-{{3 D}\over2}+s_{_2})$, respectively.
In addition, the residue of the simple pole of $\Gamma({ D\over2}-1-s_{_2})$ and that of  $\Gamma(4-D+s_{_2})$ induce
\begin{eqnarray}
&&C_{_{[52]}}^{(2)}=C_{_{[52]}}^{(5)}=-{im_{_5}^2\over(4\pi)^6}\Big({4\pi\Lambda_{_{\rm RE}}^2\over m_{_5}^2}\Big)^{6-{3D\over2}}
{\pi^2\Gamma(3-D)\over\sin^2{\pi D\over2}}\;,
\label{GKZ39c-5}
\end{eqnarray}
the residue of the simple pole of $\Gamma(D-3-s_{_2})$ and that of $\Gamma(2-{ D\over2}+s_{_2})$ induce
\begin{eqnarray}
&&C_{_{[52]}}^{(4)}=C_{_{[52]}}^{(6)}={im_{_5}^2\over(4\pi)^6}\Big({4\pi\Lambda_{_{\rm RE}}^2\over m_{_5}^2}\Big)^{6-{3D\over2}}
{\pi^2\Gamma(3-D)\over\sin^2{\pi D\over2}}\;,
\label{GKZ39c-6}
\end{eqnarray}
and $C_{_{[52]}}^{(3)}=C_{_{[52]}}^{(8)}=0$ simultaneously.
A conclusion for the case $m_{_3}\neq0$, $m_{_5}\neq0$, $m_{_1}=m_{_2}=m_{_4}=0$ is analogous
to that of $m_{_2}\neq0$, $m_{_5}\neq0$, $m_{_1}=m_{_3}=m_{_4}=0$.

\section{Conclusions\label{sec-con}}
\indent\indent
Using Miller's transformation, we derive GKZ hypergeometric systems of
Feynman integrals of the three-loop vacuum diagrams with arbitrary masses via Mellin-Barnes representation.
The dimension of dual space of the GKZ system equals the number of independent
dimensionless ratios among the virtual mass squared.
In the neighborhoods of origin including infinity, we can obtain analytical hypergeometric series
solutions of the three-loop vacuum integrals  through GKZ hypergeometric systems.
In certain nonempty intersections of corresponding convergent
regions of those hypergeometric series, the three-loop vacuum integrals can be formulated as a linear combination of those hypergeometric functions of corresponding fundamental solution system.
The combination coefficients are determined by
the vacuum  integral at some ordinary points or regular singularities, or the Mellin-Barnes representation of the vacuum  Feynman integral.

Using GKZ hypergeometric systems on general manifold, here we obtain the analytical hypergeometric solutions in the
neighborhoods of origin including infinity. In order to derive the fundamental solutions
in neighborhoods of all possible regular singularities, we can embed the general
vacuum integrals in corresponding Grassmannians manifold~\cite{Grassmannians} through their parametrization.
To efficiently derive the fundamental solution systems, next we will embed the general
three-loop vacuum integrals into the subvarieties of Grassmannians using the $\alpha$-parametric representation.

\begin{acknowledgments}
\indent\indent
The work has been supported by the National Natural Science Foundation of China (NNSFC) with Grants No. 12075074, No. 12235008, No. 11535002, No. 11705045, Hebei Natural Science Foundation for Distinguished Young Scholars with Grant No. A2022201017, Natural Science Foundation of Guangxi Autonomous Region with Grant No. 2022GXNSFDA035068, and the youth top-notch talent support program of the Hebei Province.
\end{acknowledgments}

\appendix

\section{The hypergeometric series solutions of the integer lattice ${\bf B}_{_{\tilde{1}34}}$\label{app1}}
\indent\indent
According the basis of integer lattice  $\mathbf{B}_{_{\tilde{1}34}}$,
one can construct eight GKZ hypergeometric series solutions in parameter space through choosing the sets of column indices $I_{_i}\subset [1,8]\;(i=1,\cdots,8)$.
\begin{itemize}
\item $I_{_1}=[1,2,6,7,8]$, i.e.
the implement $J_{_1}=[1,8]\setminus I_{_1}=[3,4,5]$.
The choice implies the exponent numbers $\alpha_{_3}=\alpha_{_4}=\alpha_{_5}=0$, and
\begin{eqnarray}
\alpha_{_1}=-\frac{D}{2},\;\alpha_{_2}=-1,\;
\alpha_{_6}=\alpha_{_7}=\alpha_{_8}=\frac{D}{2}-1\;.
\label{3loop-S21-1}
\end{eqnarray}
The corresponding hypergeometric series solution is written as
\begin{eqnarray}
&&\Phi_{_{[\tilde{1}34]}}^{(1)}(\alpha,z)=
y_{_1}^{{D\over2}-1}y_{_2}^{{D\over2}-1}y_{_3}^{{D\over2}-1}\sum\limits_{n_{_1}=0}^\infty
\sum\limits_{n_{_2}=0}^\infty\sum\limits_{n_{_3}=0}^\infty
{c_{_{[\tilde{1}34]}}^{(1)}(\alpha,{\bf n})}
\Big({y_{_3}}\Big)^{n_{_1}}
\Big({y_{_1}}\Big)^{n_{_2}}
\Big({y_{_2}}\Big)^{n_{_3}}\;,
\label{3loop-S21-2}
\end{eqnarray}
with the coefficient is
\begin{eqnarray}
&&c_{_{[\tilde{1}34]}}^{(1)}(\alpha,{\bf n})=
\frac{\Gamma({D\over2}+n_{_1}+n_{_2}+n_{_3}) \Gamma(1+n_{_1}+n_{_2}+n_{_3})}
{n_{_1}!n_{_2}!n_{_3}!\Gamma({D\over2}+n_{_1})\Gamma({D\over2}+n_{_2})\Gamma({D\over2}+n_{_3})}\;.
\label{3loop-S21-3}
\end{eqnarray}

\item $I_{_2}=[1,2,5,6,7]$, i.e.
the implement $J_{_2}=[1,8]\setminus I_{_2}=[3,4,8]$.
The choice implies the exponent numbers $\alpha_{_3}=\alpha_{_4}=\alpha_{_8}=0$, and
\begin{eqnarray}
\alpha_{_1}=-1,\;\alpha_{_2}=\frac{D}{2}-2,\;\alpha_{_5}=1-\frac{D}{2},\;
\alpha_{_6}=\alpha_{_7}=\frac{D}{2}-1\;.
\label{3loop-S22-1}
\end{eqnarray}
The corresponding hypergeometric series solution is written as
\begin{eqnarray}
&&\Phi_{_{[\tilde{1}34]}}^{(2)}(\alpha,z)=
y_{_1}^{{D\over2}-1}y_{_2}^{{D\over2}-1}\sum\limits_{n_{_1}=0}^\infty
\sum\limits_{n_{_2}=0}^\infty\sum\limits_{n_{_3}=0}^\infty
{c_{_{[\tilde{1}34]}}^{(2)}(\alpha,{\bf n})}
\Big({y_{_3}}\Big)^{n_{_1}}
\Big({y_{_1}}\Big)^{n_{_2}}
\Big({y_{_2}}\Big)^{n_{_3}}\;,
\label{3loop-S22-2}
\end{eqnarray}
with the coefficient is
\begin{eqnarray}
&&c_{_{[\tilde{1}34]}}^{(2)}(\alpha,{\bf n})=
\frac{\Gamma(2-{D\over2}+n_{_1}+n_{_2}+n_{_3}) \Gamma(1+n_{_1}+n_{_2}+n_{_3})}
{n_{_1}!n_{_2}!n_{_3}!\Gamma(2-{D\over2}+n_{_1})\Gamma({D\over2}+n_{_2})\Gamma({D\over2}+n_{_3})}\;.
\label{3loop-S22-3}
\end{eqnarray}

\item $I_{_3}=[1,2,4,6,8]$, i.e.
the implement $J_{_3}=[1,8]\setminus I_{_3}=[3,5,7]$.
The choice implies the exponent numbers $\alpha_{_3}=\alpha_{_5}=\alpha_{_7}=0$, and
\begin{eqnarray}
\alpha_{_1}=-1,\;\alpha_{_2}=\frac{D}{2}-2,\;\alpha_{_4}=1-\frac{D}{2},\;
\alpha_{_6}=\alpha_{_8}=\frac{D}{2}-1\;.
\label{3loop-S23-1}
\end{eqnarray}
The corresponding hypergeometric series solution is written as
\begin{eqnarray}
&&\Phi_{_{[\tilde{1}34]}}^{(3)}(\alpha,z)=
y_{_1}^{{D\over2}-1}y_{_3}^{{D\over2}-1}\sum\limits_{n_{_1}=0}^\infty
\sum\limits_{n_{_2}=0}^\infty\sum\limits_{n_{_3}=0}^\infty
{c_{_{[\tilde{1}34]}}^{(3)}(\alpha,{\bf n})}
\Big({y_{_3}}\Big)^{n_{_1}}
\Big({y_{_1}}\Big)^{n_{_2}}
\Big({y_{_2}}\Big)^{n_{_3}}\;,
\label{3loop-S23-2}
\end{eqnarray}
with the coefficient is
\begin{eqnarray}
&&c_{_{[\tilde{1}34]}}^{(3)}(\alpha,{\bf n})=
\frac{\Gamma(2-{D\over2}+n_{_1}+n_{_2}+n_{_3}) \Gamma(1+n_{_1}+n_{_2}+n_{_3})}
{n_{_1}!n_{_2}!n_{_3}!\Gamma({D\over2}+n_{_1})\Gamma({D\over2}+n_{_2})\Gamma(2-{D\over2}+n_{_3})}\;.
\label{3loop-S23-3}
\end{eqnarray}

\item $I_{_4}=[1,2,4,5,6]$, i.e.
the implement $J_{_4}=[1,8]\setminus I_{_4}=[3,7,8]$.
The choice implies the exponent numbers $\alpha_{_3}=\alpha_{_7}=\alpha_{_8}=0$, and
\begin{eqnarray}
\alpha_{_1}=\frac{D}{2}-2,\;\alpha_{_2}=D-3,\;\alpha_{_4}=\alpha_{_5}=1-\frac{D}{2},\;
\alpha_{_6}=\frac{D}{2}-1\;.
\label{3loop-S24-1}
\end{eqnarray}
The corresponding hypergeometric series solution is written as
\begin{eqnarray}
&&\Phi_{_{[\tilde{1}34]}}^{(4)}(\alpha,z)=
y_{_1}^{{D\over2}-1}\sum\limits_{n_{_1}=0}^\infty
\sum\limits_{n_{_2}=0}^\infty\sum\limits_{n_{_3}=0}^\infty
{c_{_{[\tilde{1}34]}}^{(4)}(\alpha,{\bf n})}
\Big({y_{_3}}\Big)^{n_{_1}}
\Big({y_{_1}}\Big)^{n_{_2}}
\Big({y_{_2}}\Big)^{n_{_3}}\;,
\label{3loop-S24-2}
\end{eqnarray}
with the coefficient is
\begin{eqnarray}
&&c_{_{[\tilde{1}34]}}^{(4)}(\alpha,{\bf n})=
\frac{\Gamma(2-{D\over2}+n_{_1}+n_{_2}+n_{_3}) \Gamma(3-D+n_{_1}+n_{_2}+n_{_3})}
{n_{_1}!n_{_2}!n_{_3}!\Gamma(2-{D\over2}+n_{_1})\Gamma({D\over2}+n_{_2})\Gamma(2-{D\over2}+n_{_3})}\;.
\label{3loop-S24-3}
\end{eqnarray}

\item $I_{_5}=[1,2,3,7,8]$, i.e.
the implement $J_{_5}=[1,8]\setminus I_{_5}=[4,5,6]$.
The choice implies the exponent numbers $\alpha_{_4}=\alpha_{_5}=\alpha_{_6}=0$, and
\begin{eqnarray}
\alpha_{_1}=-1,\;\alpha_{_2}=\frac{D}{2}-2,\;\alpha_{_3}=1-\frac{D}{2},\;
\alpha_{_7}=\alpha_{_8}=\frac{D}{2}-1\;.
\label{3loop-S25-1}
\end{eqnarray}
The corresponding hypergeometric series solution is written as
\begin{eqnarray}
&&\Phi_{_{[\tilde{1}34]}}^{(5)}(\alpha,z)=
y_{_2}^{{D\over2}-1}y_{_3}^{{D\over2}-1}\sum\limits_{n_{_1}=0}^\infty
\sum\limits_{n_{_2}=0}^\infty\sum\limits_{n_{_3}=0}^\infty
{c_{_{[\tilde{1}34]}}^{(5)}(\alpha,{\bf n})}
\Big({y_{_3}}\Big)^{n_{_1}}
\Big({y_{_1}}\Big)^{n_{_2}}
\Big({y_{_2}}\Big)^{n_{_3}}\;,
\label{3loop-S25-2}
\end{eqnarray}
with the coefficient is
\begin{eqnarray}
&&c_{_{[\tilde{1}34]}}^{(5)}(\alpha,{\bf n})=
\frac{\Gamma(2-{D\over2}+n_{_1}+n_{_2}+n_{_3}) \Gamma(1+n_{_1}+n_{_2}+n_{_3})}
{n_{_1}!n_{_2}!n_{_3}!\Gamma({D\over2}+n_{_1})\Gamma(2-{D\over2}+n_{_2})\Gamma({D\over2}+n_{_3})}\;.
\label{3loop-S25-3}
\end{eqnarray}

\item $I_{_6}=[1,2,3,5,7]$, i.e.
the implement $J_{_6}=[1,8]\setminus I_{_6}=[4,6,8]$.
The choice implies the exponent numbers $\alpha_{_4}=\alpha_{_6}=\alpha_{_8}=0$, and
\begin{eqnarray}
\alpha_{_1}=\frac{D}{2}-2,\;\alpha_{_2}=D-3,\;\alpha_{_3}=\alpha_{_5}=1-\frac{D}{2},\;
\alpha_{_7}=\frac{D}{2}-1\;.
\label{3loop-S26-1}
\end{eqnarray}
The corresponding hypergeometric series solution is written as
\begin{eqnarray}
&&\Phi_{_{[\tilde{1}34]}}^{(6)}(\alpha,z)=
y_{_2}^{{D\over2}-1}\sum\limits_{n_{_1}=0}^\infty
\sum\limits_{n_{_2}=0}^\infty\sum\limits_{n_{_3}=0}^\infty
{c_{_{[\tilde{1}34]}}^{(6)}(\alpha,{\bf n})}
\Big({y_{_3}}\Big)^{n_{_1}}
\Big({y_{_1}}\Big)^{n_{_2}}
\Big({y_{_2}}\Big)^{n_{_3}}\;,
\label{3loop-S26-2}
\end{eqnarray}
with the coefficient is
\begin{eqnarray}
&&c_{_{[\tilde{1}34]}}^{(6)}(\alpha,{\bf n})=
\frac{\Gamma(2-{D\over2}+n_{_1}+n_{_2}+n_{_3}) \Gamma(3-D+n_{_1}+n_{_2}+n_{_3})}
{n_{_1}!n_{_2}!n_{_3}!\Gamma(2-{D\over2}+n_{_1})\Gamma(2-{D\over2}+n_{_2})\Gamma({D\over2}+n_{_3})}\;.
\label{3loop-S26-3}
\end{eqnarray}

\item $I_{_7}=[1,2,3,4,8]$, i.e.
the implement $J_{_7}=[1,8]\setminus I_{_7}=[5,6,7]$.
The choice implies the exponent numbers $\alpha_{_5}=\alpha_{_6}=\alpha_{_7}=0$, and
\begin{eqnarray}
\alpha_{_1}=\frac{D}{2}-2,\;\alpha_{_2}=D-3,\;\alpha_{_3}=\alpha_{_4}=1-\frac{D}{2},\;
\alpha_{_8}=\frac{D}{2}-1\;.
\label{3loop-S27-1}
\end{eqnarray}
The corresponding hypergeometric series solution is written as
\begin{eqnarray}
&&\Phi_{_{[\tilde{1}34]}}^{(7)}(\alpha,z)=
y_{_3}^{{D\over2}-1}\sum\limits_{n_{_1}=0}^\infty
\sum\limits_{n_{_2}=0}^\infty\sum\limits_{n_{_3}=0}^\infty
{c_{_{[\tilde{1}34]}}^{(7)}(\alpha,{\bf n})}
\Big({y_{_3}}\Big)^{n_{_1}}
\Big({y_{_1}}\Big)^{n_{_2}}
\Big({y_{_2}}\Big)^{n_{_3}}\;,
\label{3loop-S27-2}
\end{eqnarray}
with the coefficient is
\begin{eqnarray}
&&c_{_{[\tilde{1}34]}}^{(7)}(\alpha,{\bf n})=
\frac{\Gamma(2-{D\over2}+n_{_1}+n_{_2}+n_{_3}) \Gamma(3-D+n_{_1}+n_{_2}+n_{_3})}
{n_{_1}!n_{_2}!n_{_3}!\Gamma({D\over2}+n_{_1})\Gamma(2-{D\over2}+n_{_2})\Gamma(2-{D\over2}+n_{_3})}\;.
\label{3loop-S27-3}
\end{eqnarray}

\item $I_{_8}=[1,2,3,4,5]$, i.e.
the implement $J_{_8}=[1,8]\setminus I_{_8}=[6,7,8]$.
The choice implies the exponent numbers $\alpha_{_6}=\alpha_{_7}=\alpha_{_8}=0$, and
\begin{eqnarray}
\alpha_{_1}=D-3,\;\alpha_{_2}=\frac{3D}{2}-4,\;
\alpha_{_3}=\alpha_{_4}=\alpha_{_5}=1-\frac{D}{2}\;.
\label{3loop-S28-1}
\end{eqnarray}
The corresponding hypergeometric series solution is written as
\begin{eqnarray}
&&\Phi_{_{[\tilde{1}34]}}^{(8)}(\alpha,z)=
\sum\limits_{n_{_1}=0}^\infty
\sum\limits_{n_{_2}=0}^\infty\sum\limits_{n_{_3}=0}^\infty
{c_{_{[\tilde{1}34]}}^{(8)}(\alpha,{\bf n})}
\Big({y_{_3}}\Big)^{n_{_1}}
\Big({y_{_1}}\Big)^{n_{_2}}
\Big({y_{_2}}\Big)^{n_{_3}}\;,
\label{3loop-S28-2}
\end{eqnarray}
with the coefficient is
\begin{eqnarray}
&&c_{_{[\tilde{1}34]}}^{(8)}(\alpha,{\bf n})=
\frac{\Gamma(4-{3D\over2}+n_{_1}+n_{_2}+n_{_3}) \Gamma(3-D+n_{_1}+n_{_2}+n_{_3})}
{n_{_1}!n_{_2}!n_{_3}!\Gamma(2-{D\over2}+n_{_1})\Gamma(2-{D\over2}+n_{_2})\Gamma(2-{D\over2}+n_{_3})}\;.
\label{3loop-S28-3}
\end{eqnarray}

\end{itemize}

\section{The hypergeometric series solutions of the integer lattice ${\bf B}_{_{1\tilde{3}4}}$\label{app2}}
\indent\indent
According the basis of integer lattice  $\mathbf{B}_{_{1\tilde{3}4}}$,
one can construct eight GKZ hypergeometric series solutions in parameter space through choosing the sets of column indices $I_{_i}\subset [1,8]\;(i=1,\cdots,8)$.
\begin{itemize}
\item $I_{_1}=[2,3,6,7,8]$, i.e.
the implement $J_{_1}=[1,8]\setminus I_{_1}=[1,4,5]$.
The choice implies the exponent numbers $\alpha_{_1}=\alpha_{_4}=\alpha_{_5}=0$, and
\begin{eqnarray}
\alpha_{_2}=\frac{D}{2}-1,\;\alpha_{_3}=-\frac{D}{2},\;\alpha_{_6}=-1,\;
\alpha_{_7}=\alpha_{_8}=\frac{D}{2}-1\;.
\label{3loop-S31-1}
\end{eqnarray}
The corresponding hypergeometric series solution is written as
\begin{eqnarray}
&&\Phi_{_{[1\tilde{3}4]}}^{(1)}(\alpha,z)=
y_{_1}^{-1}y_{_2}^{{D\over2}-1}y_{_3}^{{D\over2}-1}\sum\limits_{n_{_1}=0}^\infty
\sum\limits_{n_{_2}=0}^\infty\sum\limits_{n_{_3}=0}^\infty
{c_{_{[1\tilde{3}4]}}^{(1)}(\alpha,{\bf n})}
\Big({1\over y_{_1}}\Big)^{n_{_1}}
\Big({y_{_3}\over y_{_1}}\Big)^{n_{_2}}
\Big({y_{_2}\over y_{_1}}\Big)^{n_{_3}}\;,
\label{3loop-S31-2}
\end{eqnarray}
with the coefficient is
\begin{eqnarray}
&&c_{_{[1\tilde{3}4]}}^{(1)}(\alpha,{\bf n})=
\frac{\Gamma({D\over2}+n_{_1}+n_{_2}+n_{_3}) \Gamma(1+n_{_1}+n_{_2}+n_{_3})}
{n_{_1}!n_{_2}!n_{_3}!\Gamma({D\over2}+n_{_1})\Gamma({D\over2}+n_{_2})\Gamma({D\over2}+n_{_3})}\;.
\label{3loop-S31-3}
\end{eqnarray}

\item $I_{_2}=[2,3,5,6,7]$, i.e.
the implement $J_{_2}=[1,8]\setminus I_{_2}=[1,4,8]$.
The choice implies the exponent numbers $\alpha_{_1}=\alpha_{_4}=\alpha_{_8}=0$, and
\begin{eqnarray}
\alpha_{_2}=\alpha_{_7}=\frac{D}{2}-1,\;\alpha_{_3}=-1,\;\alpha_{_5}=1-\frac{D}{2},\;
\alpha_{_6}=\frac{D}{2}-2\;.
\label{3loop-S32-1}
\end{eqnarray}
The corresponding hypergeometric series solution is written as
\begin{eqnarray}
&&\Phi_{_{[1\tilde{3}4]}}^{(2)}(\alpha,z)=
y_{_1}^{{D\over2}-2}y_{_2}^{{D\over2}-1}\sum\limits_{n_{_1}=0}^\infty
\sum\limits_{n_{_2}=0}^\infty\sum\limits_{n_{_3}=0}^\infty
{c_{_{[1\tilde{3}4]}}^{(2)}(\alpha,{\bf n})}
\Big({1\over y_{_1}}\Big)^{n_{_1}}
\Big({y_{_3}\over y_{_1}}\Big)^{n_{_2}}
\Big({y_{_2}\over y_{_1}}\Big)^{n_{_3}}\;,
\label{3loop-S32-2}
\end{eqnarray}
with the coefficient is
\begin{eqnarray}
&&c_{_{[1\tilde{3}4]}}^{(2)}(\alpha,{\bf n})=
\frac{\Gamma(2-{D\over2}+n_{_1}+n_{_2}+n_{_3}) \Gamma(1+n_{_1}+n_{_2}+n_{_3})}
{n_{_1}!n_{_2}!n_{_3}!\Gamma({D\over2}+n_{_1})\Gamma(2-{D\over2}+n_{_2})\Gamma({D\over2}+n_{_3})}\;.
\label{3loop-S32-3}
\end{eqnarray}

\item $I_{_3}=[2,3,4,6,8]$, i.e.
the implement $J_{_3}=[1,8]\setminus I_{_3}=[1,5,7]$.
The choice implies the exponent numbers $\alpha_{_1}=\alpha_{_5}=\alpha_{_7}=0$, and
\begin{eqnarray}
\alpha_{_2}=\alpha_{_8}=\frac{D}{2}-1,\;\alpha_{_3}=-1,\;\alpha_{_4}=1-\frac{D}{2},\;
\alpha_{_6}=\frac{D}{2}-2\;.
\label{3loop-S33-1}
\end{eqnarray}
The corresponding hypergeometric series solution is written as
\begin{eqnarray}
&&\Phi_{_{[1\tilde{3}4]}}^{(3)}(\alpha,z)=
y_{_1}^{{D\over2}-2}y_{_3}^{{D\over2}-1}\sum\limits_{n_{_1}=0}^\infty
\sum\limits_{n_{_2}=0}^\infty\sum\limits_{n_{_3}=0}^\infty
{c_{_{[1\tilde{3}4]}}^{(3)}(\alpha,{\bf n})}
\Big({1\over y_{_1}}\Big)^{n_{_1}}
\Big({y_{_3}\over y_{_1}}\Big)^{n_{_2}}
\Big({y_{_2}\over y_{_1}}\Big)^{n_{_3}}\;,
\label{3loop-S33-2}
\end{eqnarray}
with the coefficient is
\begin{eqnarray}
&&c_{_{[1\tilde{3}4]}}^{(3)}(\alpha,{\bf n})=
\frac{\Gamma(2-{D\over2}+n_{_1}+n_{_2}+n_{_3}) \Gamma(1+n_{_1}+n_{_2}+n_{_3})}
{n_{_1}!n_{_2}!n_{_3}!\Gamma({D\over2}+n_{_1})\Gamma({D\over2}+n_{_2})\Gamma(2-{D\over2}+n_{_3})}\;.
\label{3loop-S33-3}
\end{eqnarray}

\item $I_{_4}=[2,3,4,5,6]$, i.e.
the implement $J_{_4}=[1,8]\setminus I_{_4}=[1,7,8]$.
The choice implies the exponent numbers $\alpha_{_1}=\alpha_{_7}=\alpha_{_8}=0$, and
\begin{eqnarray}
\alpha_{_2}=\frac{D}{2}-1,\;\alpha_{_3}=\frac{D}{2}-2,\;\alpha_{_4}=\alpha_{_5}=1-\frac{D}{2},\;
\alpha_{_6}=D-3\;.
\label{3loop-S34-1}
\end{eqnarray}
The corresponding hypergeometric series solution is written as
\begin{eqnarray}
&&\Phi_{_{[1\tilde{3}4]}}^{(4)}(\alpha,z)=
y_{_1}^{D-3}\sum\limits_{n_{_1}=0}^\infty
\sum\limits_{n_{_2}=0}^\infty\sum\limits_{n_{_3}=0}^\infty
{c_{_{[1\tilde{3}4]}}^{(4)}(\alpha,{\bf n})}
\Big({1\over y_{_1}}\Big)^{n_{_1}}
\Big({y_{_3}\over y_{_1}}\Big)^{n_{_2}}
\Big({y_{_2}\over y_{_1}}\Big)^{n_{_3}}\;,
\label{3loop-S34-2}
\end{eqnarray}
with the coefficient is
\begin{eqnarray}
&&c_{_{[1\tilde{3}4]}}^{(4)}(\alpha,{\bf n})=
\frac{\Gamma(2-{D\over2}+n_{_1}+n_{_2}+n_{_3}) \Gamma(3-D+n_{_1}+n_{_2}+n_{_3})}
{n_{_1}!n_{_2}!n_{_3}!\Gamma({D\over2}+n_{_1})\Gamma(2-{D\over2}+n_{_2})\Gamma(2-{D\over2}+n_{_3})}\;.
\label{3loop-S34-3}
\end{eqnarray}

\item $I_{_5}=[1,3,6,7,8]$, i.e.
the implement $J_{_5}=[1,8]\setminus I_{_5}=[2,4,5]$.
The choice implies the exponent numbers $\alpha_{_2}=\alpha_{_4}=\alpha_{_5}=0$, and
\begin{eqnarray}
\alpha_{_1}=1-\frac{D}{2},\;\alpha_{_3}=-1,\;\alpha_{_6}=\frac{D}{2}-2,\;
\alpha_{_7}=\alpha_{_8}=\frac{D}{2}-1\;.
\label{3loop-S35-1}
\end{eqnarray}
The corresponding hypergeometric series solution is written as
\begin{eqnarray}
&&\hspace{-0.9cm}\Phi_{_{[1\tilde{3}4]}}^{(5)}(\alpha,z)=
y_{_1}^{\frac{D}{2}-2}y_{_2}^{{D\over2}-1}y_{_3}^{{D\over2}-1}\sum\limits_{n_{_1}=0}^\infty
\sum\limits_{n_{_2}=0}^\infty\sum\limits_{n_{_3}=0}^\infty
{c_{_{[1\tilde{3}4]}}^{(5)}(\alpha,{\bf n})}
\Big({1\over y_{_1}}\Big)^{n_{_1}}
\Big({y_{_3}\over y_{_1}}\Big)^{n_{_2}}
\Big({y_{_2}\over y_{_1}}\Big)^{n_{_3}}\;,
\label{3loop-S35-2}
\end{eqnarray}
with the coefficient is
\begin{eqnarray}
&&c_{_{[1\tilde{3}4]}}^{(5)}(\alpha,{\bf n})=
\frac{\Gamma(2-{D\over2}+n_{_1}+n_{_2}+n_{_3}) \Gamma(1+n_{_1}+n_{_2}+n_{_3})}
{n_{_1}!n_{_2}!n_{_3}!\Gamma(2-{D\over2}+n_{_1})\Gamma({D\over2}+n_{_2})\Gamma({D\over2}+n_{_3})}\;.
\label{3loop-S35-3}
\end{eqnarray}

\item $I_{_6}=[1,3,5,6,7]$, i.e.
the implement $J_{_6}=[1,8]\setminus I_{_6}=[2,4,8]$.
The choice implies the exponent numbers $\alpha_{_2}=\alpha_{_4}=\alpha_{_8}=0$, and
\begin{eqnarray}
\alpha_{_1}=\alpha_{_5}=1-\frac{D}{2},\;\alpha_{_3}=\frac{D}{2}-2,\;
\alpha_{_6}=D-3,\;\alpha_{_7}=\frac{D}{2}-1\;.
\label{3loop-S36-1}
\end{eqnarray}
The corresponding hypergeometric series solution is written as
\begin{eqnarray}
&&\Phi_{_{[1\tilde{3}4]}}^{(6)}(\alpha,z)=
y_{_1}^{D-3}y_{_2}^{{D\over2}-1}\sum\limits_{n_{_1}=0}^\infty
\sum\limits_{n_{_2}=0}^\infty\sum\limits_{n_{_3}=0}^\infty
{c_{_{[1\tilde{3}4]}}^{(6)}(\alpha,{\bf n})}
\Big({1\over y_{_1}}\Big)^{n_{_1}}
\Big({y_{_3}\over y_{_1}}\Big)^{n_{_2}}
\Big({y_{_2}\over y_{_1}}\Big)^{n_{_3}}\;,
\label{3loop-S36-2}
\end{eqnarray}
with the coefficient is
\begin{eqnarray}
&&c_{_{[1\tilde{3}4]}}^{(6)}(\alpha,{\bf n})=
\frac{\Gamma(2-{D\over2}+n_{_1}+n_{_2}+n_{_3}) \Gamma(3-D+n_{_1}+n_{_2}+n_{_3})}
{n_{_1}!n_{_2}!n_{_3}!\Gamma(2-{D\over2}+n_{_1})\Gamma(2-{D\over2}+n_{_2})\Gamma({D\over2}+n_{_3})}\;.
\label{3loop-S36-3}
\end{eqnarray}

\item $I_{_7}=[1,3,4,6,8]$, i.e.
the implement $J_{_7}=[1,8]\setminus I_{_7}=[2,5,7]$.
The choice implies the exponent numbers $\alpha_{_2}=\alpha_{_5}=\alpha_{_7}=0$, and
\begin{eqnarray}
\alpha_{_1}=\alpha_{_4}=1-\frac{D}{2},\;\alpha_{_3}=\frac{D}{2}-2,\;
\alpha_{_6}=D-3,\;\alpha_{_8}=\frac{D}{2}-1\;.
\label{3loop-S37-1}
\end{eqnarray}
The corresponding hypergeometric series solution is written as
\begin{eqnarray}
&&\Phi_{_{[1\tilde{3}4]}}^{(7)}(\alpha,z)=
y_{_1}^{D-3}y_{_3}^{{D\over2}-1}\sum\limits_{n_{_1}=0}^\infty
\sum\limits_{n_{_2}=0}^\infty\sum\limits_{n_{_3}=0}^\infty
{c_{_{[1\tilde{3}4]}}^{(7)}(\alpha,{\bf n})}
\Big({1\over y_{_1}}\Big)^{n_{_1}}
\Big({y_{_3}\over y_{_1}}\Big)^{n_{_2}}
\Big({y_{_2}\over y_{_1}}\Big)^{n_{_3}}\;,
\label{3loop-S37-2}
\end{eqnarray}
with the coefficient is
\begin{eqnarray}
&&c_{_{[1\tilde{3}4]}}^{(7)}(\alpha,{\bf n})=
\frac{\Gamma(2-{D\over2}+n_{_1}+n_{_2}+n_{_3}) \Gamma(3-D+n_{_1}+n_{_2}+n_{_3})}
{n_{_1}!n_{_2}!n_{_3}!\Gamma(2-{D\over2}+n_{_1})\Gamma({D\over2}+n_{_2})\Gamma(2-{D\over2}+n_{_3})}\;.
\label{3loop-S37-3}
\end{eqnarray}

\item $I_{_8}=[1,3,4,5,6]$, i.e.
the implement $J_{_8}=[1,8]\setminus I_{_8}=[2,7,8]$.
The choice implies the exponent numbers $\alpha_{_2}=\alpha_{_7}=\alpha_{_8}=0$, and
\begin{eqnarray}
\alpha_{_1}=\alpha_{_4}=\alpha_{_5}=1-\frac{D}{2},\;\alpha_{_3}=D-3,\;
\alpha_{_6}=\frac{3D}{2}-4\;.
\label{3loop-S38-1}
\end{eqnarray}
The corresponding hypergeometric series solution is written as
\begin{eqnarray}
&&\Phi_{_{[1\tilde{3}4]}}^{(8)}(\alpha,z)=
y_{_1}^{\frac{3D}{2}-4}\sum\limits_{n_{_1}=0}^\infty
\sum\limits_{n_{_2}=0}^\infty\sum\limits_{n_{_3}=0}^\infty
{c_{_{[1\tilde{3}4]}}^{(8)}(\alpha,{\bf n})}
\Big({1\over y_{_1}}\Big)^{n_{_1}}
\Big({y_{_3}\over y_{_1}}\Big)^{n_{_2}}
\Big({y_{_2}\over y_{_1}}\Big)^{n_{_3}}\;,
\label{3loop-S38-2}
\end{eqnarray}
with the coefficient is
\begin{eqnarray}
&&c_{_{[1\tilde{3}4]}}^{(8)}(\alpha,{\bf n})=
\frac{\Gamma(4-{3D\over2}+n_{_1}+n_{_2}+n_{_3}) \Gamma(3-D+n_{_1}+n_{_2}+n_{_3})}
{n_{_1}!n_{_2}!n_{_3}!\Gamma(2-{D\over2}+n_{_1})\Gamma(2-{D\over2}+n_{_2})\Gamma(2-{D\over2}+n_{_3})}\;.
\label{3loop-S38-3}
\end{eqnarray}

\end{itemize}

\section{The hypergeometric series solutions of the integer lattice ${\bf B}_{_{13\tilde{4}}}$\label{app3}}
\indent\indent
According the basis of integer lattice  $\mathbf{B}_{_{13\tilde{4}}}$,
one can construct eight GKZ hypergeometric series solutions in parameter space through choosing the sets of column indices $I_{_i}\subset [1,8]\;(i=1,\cdots,8)$.
\begin{itemize}
\item $I_{_1}=[2,4,6,7,8]$, i.e.
the implement $J_{_1}=[1,8]\setminus I_{_1}=[1,3,5]$.
The choice implies the exponent numbers $\alpha_{_1}=\alpha_{_3}=\alpha_{_5}=0$, and
\begin{eqnarray}
\alpha_{_2}=\frac{D}{2}-1,\;\alpha_{_4}=-\frac{D}{2},\;\alpha_{_7}=-1,\;
\alpha_{_6}=\alpha_{_8}=\frac{D}{2}-1\;.
\label{3loop-S41-1}
\end{eqnarray}
The corresponding hypergeometric series solution is written as
\begin{eqnarray}
&&\Phi_{_{[13\tilde{4}]}}^{(1)}(\alpha,z)=
y_{_1}^{{D\over2}-1}y_{_2}^{-1}y_{_3}^{{D\over2}-1}\sum\limits_{n_{_1}=0}^\infty
\sum\limits_{n_{_2}=0}^\infty\sum\limits_{n_{_3}=0}^\infty
{c_{_{[13\tilde{4}]}}^{(1)}(\alpha,{\bf n})}
\Big({1\over y_{_2}}\Big)^{n_{_1}}
\Big({y_{_1}\over y_{_2}}\Big)^{n_{_2}}
\Big({y_{_3}\over y_{_2}}\Big)^{n_{_3}}\;,
\label{3loop-S41-2}
\end{eqnarray}
with the coefficient is
\begin{eqnarray}
&&c_{_{[13\tilde{4}]}}^{(1)}(\alpha,{\bf n})=
\frac{\Gamma({D\over2}+n_{_1}+n_{_2}+n_{_3}) \Gamma(1+n_{_1}+n_{_2}+n_{_3})}
{n_{_1}!n_{_2}!n_{_3}!\Gamma({D\over2}+n_{_1})\Gamma({D\over2}+n_{_2})\Gamma({D\over2}+n_{_3})}\;.
\label{3loop-S41-3}
\end{eqnarray}

\item $I_{_2}=[2,4,5,6,7]$, i.e.
the implement $J_{_2}=[1,8]\setminus I_{_2}=[1,3,8]$.
The choice implies the exponent numbers $\alpha_{_1}=\alpha_{_3}=\alpha_{_8}=0$, and
\begin{eqnarray}
\alpha_{_2}=\alpha_{_6}=\frac{D}{2}-1,\;\alpha_{_4}=-1,\;\alpha_{_5}=1-\frac{D}{2},\;
\alpha_{_7}=\frac{D}{2}-2\;.
\label{3loop-S42-1}
\end{eqnarray}
The corresponding hypergeometric series solution is written as
\begin{eqnarray}
&&\Phi_{_{[13\tilde{4}]}}^{(2)}(\alpha,z)=
y_{_1}^{{D\over2}-1}y_{_2}^{{D\over2}-2}\sum\limits_{n_{_1}=0}^\infty
\sum\limits_{n_{_2}=0}^\infty\sum\limits_{n_{_3}=0}^\infty
{c_{_{[13\tilde{4}]}}^{(2)}(\alpha,{\bf n})}
\Big({1\over y_{_2}}\Big)^{n_{_1}}
\Big({y_{_1}\over y_{_2}}\Big)^{n_{_2}}
\Big({y_{_3}\over y_{_2}}\Big)^{n_{_3}}\;,
\label{3loop-S42-2}
\end{eqnarray}
with the coefficient is
\begin{eqnarray}
&&c_{_{[13\tilde{4}]}}^{(2)}(\alpha,{\bf n})=
\frac{\Gamma(2-{D\over2}+n_{_1}+n_{_2}+n_{_3}) \Gamma(1+n_{_1}+n_{_2}+n_{_3})}
{n_{_1}!n_{_2}!n_{_3}!\Gamma({D\over2}+n_{_1})\Gamma({D\over2}+n_{_2})\Gamma(2-{D\over2}+n_{_3})}\;.
\label{3loop-S42-3}
\end{eqnarray}

\item $I_{_3}=[2,3,4,7,8]$, i.e.
the implement $J_{_3}=[1,8]\setminus I_{_3}=[1,5,6]$.
The choice implies the exponent numbers $\alpha_{_1}=\alpha_{_5}=\alpha_{_6}=0$, and
\begin{eqnarray}
\alpha_{_2}=\alpha_{_8}=\frac{D}{2}-1,\;\alpha_{_3}=1-\frac{D}{2},\;\alpha_{_4}=-1,\;
\alpha_{_7}=\frac{D}{2}-2\;.
\label{3loop-S43-1}
\end{eqnarray}
The corresponding hypergeometric series solution is written as
\begin{eqnarray}
&&\Phi_{_{[13\tilde{4}]}}^{(3)}(\alpha,z)=
y_{_2}^{{D\over2}-2}y_{_3}^{{D\over2}-1}\sum\limits_{n_{_1}=0}^\infty
\sum\limits_{n_{_2}=0}^\infty\sum\limits_{n_{_3}=0}^\infty
{c_{_{[13\tilde{4}]}}^{(3)}(\alpha,{\bf n})}
\Big({1\over y_{_2}}\Big)^{n_{_1}}
\Big({y_{_1}\over y_{_2}}\Big)^{n_{_2}}
\Big({y_{_3}\over y_{_2}}\Big)^{n_{_3}}\;,
\label{3loop-S43-2}
\end{eqnarray}
with the coefficient is
\begin{eqnarray}
&&c_{_{[13\tilde{4}]}}^{(3)}(\alpha,{\bf n})=
\frac{\Gamma(2-{D\over2}+n_{_1}+n_{_2}+n_{_3}) \Gamma(1+n_{_1}+n_{_2}+n_{_3})}
{n_{_1}!n_{_2}!n_{_3}!\Gamma({D\over2}+n_{_1})\Gamma(2-{D\over2}+n_{_2})\Gamma({D\over2}+n_{_3})}\;.
\label{3loop-S43-3}
\end{eqnarray}

\item $I_{_4}=[2,3,4,5,7]$, i.e.
the implement $J_{_4}=[1,8]\setminus I_{_4}=[1,6,8]$.
The choice implies the exponent numbers $\alpha_{_1}=\alpha_{_6}=\alpha_{_8}=0$, and
\begin{eqnarray}
\alpha_{_2}=\frac{D}{2}-1,\;\alpha_{_3}=\alpha_{_5}=1-\frac{D}{2},\;
\alpha_{_4}=\frac{D}{2}-2,\;\alpha_{_7}=D-3\;.
\label{3loop-S44-1}
\end{eqnarray}
The corresponding hypergeometric series solution is written as
\begin{eqnarray}
&&\Phi_{_{[13\tilde{4}]}}^{(4)}(\alpha,z)=
y_{_2}^{D-3}\sum\limits_{n_{_1}=0}^\infty
\sum\limits_{n_{_2}=0}^\infty\sum\limits_{n_{_3}=0}^\infty
{c_{_{[13\tilde{4}]}}^{(4)}(\alpha,{\bf n})}
\Big({1\over y_{_2}}\Big)^{n_{_1}}
\Big({y_{_1}\over y_{_2}}\Big)^{n_{_2}}
\Big({y_{_3}\over y_{_2}}\Big)^{n_{_3}}\;,
\label{3loop-S44-2}
\end{eqnarray}
with the coefficient is
\begin{eqnarray}
&&c_{_{[13\tilde{4}]}}^{(4)}(\alpha,{\bf n})=
\frac{\Gamma(2-{D\over2}+n_{_1}+n_{_2}+n_{_3}) \Gamma(3-D+n_{_1}+n_{_2}+n_{_3})}
{n_{_1}!n_{_2}!n_{_3}!\Gamma({D\over2}+n_{_1})\Gamma(2-{D\over2}+n_{_2})\Gamma(2-{D\over2}+n_{_3})}\;.
\label{3loop-S44-3}
\end{eqnarray}

\item $I_{_5}=[1,4,6,7,8]$, i.e.
the implement $J_{_5}=[1,8]\setminus I_{_5}=[2,3,5]$.
The choice implies the exponent numbers $\alpha_{_2}=\alpha_{_3}=\alpha_{_5}=0$, and
\begin{eqnarray}
\alpha_{_1}=1-\frac{D}{2},\;\alpha_{_4}=-1,\;\alpha_{_7}=\frac{D}{2}-2,\;
\alpha_{_6}=\alpha_{_8}=\frac{D}{2}-1\;.
\label{3loop-S45-1}
\end{eqnarray}
The corresponding hypergeometric series solution is written as
\begin{eqnarray}
&&\Phi_{_{[13\tilde{4}]}}^{(5)}(\alpha,z)=
y_{_1}^{{D\over2}-1}y_{_2}^{\frac{D}{2}-2}y_{_3}^{{D\over2}-1}\sum\limits_{n_{_1}=0}^\infty
\sum\limits_{n_{_2}=0}^\infty\sum\limits_{n_{_3}=0}^\infty
{c_{_{[13\tilde{4}]}}^{(5)}(\alpha,{\bf n})}
\Big({1\over y_{_2}}\Big)^{n_{_1}}
\Big({y_{_1}\over y_{_2}}\Big)^{n_{_2}}
\Big({y_{_3}\over y_{_2}}\Big)^{n_{_3}}\;,
\label{3loop-S45-2}
\end{eqnarray}
with the coefficient is
\begin{eqnarray}
&&c_{_{[13\tilde{4}]}}^{(5)}(\alpha,{\bf n})=
\frac{\Gamma(2-{D\over2}+n_{_1}+n_{_2}+n_{_3}) \Gamma(1+n_{_1}+n_{_2}+n_{_3})}
{n_{_1}!n_{_2}!n_{_3}!\Gamma(2-{D\over2}+n_{_1})\Gamma({D\over2}+n_{_2})\Gamma({D\over2}+n_{_3})}\;.
\label{3loop-S45-3}
\end{eqnarray}

\item $I_{_6}=[1,4,5,6,7]$, i.e.
the implement $J_{_6}=[1,8]\setminus I_{_6}=[2,3,8]$.
The choice implies the exponent numbers $\alpha_{_2}=\alpha_{_3}=\alpha_{_8}=0$, and
\begin{eqnarray}
\alpha_{_1}=\alpha_{_5}=1-\frac{D}{2},\;\alpha_{_4}=\frac{D}{2}-2,\;
\alpha_{_6}=\frac{D}{2}-1,\;\alpha_{_7}=D-3\;.
\label{3loop-S46-1}
\end{eqnarray}
The corresponding hypergeometric series solution is written as
\begin{eqnarray}
&&\Phi_{_{[13\tilde{4}]}}^{(6)}(\alpha,z)=
y_{_1}^{{D\over2}-1}y_{_2}^{D-3}\sum\limits_{n_{_1}=0}^\infty
\sum\limits_{n_{_2}=0}^\infty\sum\limits_{n_{_3}=0}^\infty
{c_{_{[13\tilde{4}]}}^{(6)}(\alpha,{\bf n})}
\Big({1\over y_{_2}}\Big)^{n_{_1}}
\Big({y_{_1}\over y_{_2}}\Big)^{n_{_2}}
\Big({y_{_3}\over y_{_2}}\Big)^{n_{_3}}\;,
\label{3loop-S46-2}
\end{eqnarray}
with the coefficient is
\begin{eqnarray}
&&c_{_{[13\tilde{4}]}}^{(6)}(\alpha,{\bf n})=
\frac{\Gamma(2-{D\over2}+n_{_1}+n_{_2}+n_{_3}) \Gamma(3-D+n_{_1}+n_{_2}+n_{_3})}
{n_{_1}!n_{_2}!n_{_3}!\Gamma(2-{D\over2}+n_{_1})\Gamma({D\over2}+n_{_2})\Gamma(2-{D\over2}+n_{_3})}\;.
\label{3loop-S46-3}
\end{eqnarray}

\item $I_{_7}=[1,3,4,7,8]$, i.e.
the implement $J_{_7}=[1,8]\setminus I_{_7}=[2,5,6]$.
The choice implies the exponent numbers $\alpha_{_2}=\alpha_{_5}=\alpha_{_6}=0$, and
\begin{eqnarray}
\alpha_{_1}=\alpha_{_3}=1-\frac{D}{2},\;\alpha_{_4}=\frac{D}{2}-2,\;
\alpha_{_7}=D-3,\;\alpha_{_8}=\frac{D}{2}-1\;.
\label{3loop-S47-1}
\end{eqnarray}
The corresponding hypergeometric series solution is written as
\begin{eqnarray}
&&\Phi_{_{[13\tilde{4}]}}^{(7)}(\alpha,z)=
y_{_2}^{D-3}y_{_3}^{{D\over2}-1}\sum\limits_{n_{_1}=0}^\infty
\sum\limits_{n_{_2}=0}^\infty\sum\limits_{n_{_3}=0}^\infty
{c_{_{[13\tilde{4}]}}^{(7)}(\alpha,{\bf n})}
\Big({1\over y_{_2}}\Big)^{n_{_1}}
\Big({y_{_1}\over y_{_2}}\Big)^{n_{_2}}
\Big({y_{_3}\over y_{_2}}\Big)^{n_{_3}}\;,
\label{3loop-S47-2}
\end{eqnarray}
with the coefficient is
\begin{eqnarray}
&&c_{_{[13\tilde{4}]}}^{(7)}(\alpha,{\bf n})=
\frac{\Gamma(2-{D\over2}+n_{_1}+n_{_2}+n_{_3}) \Gamma(3-D+n_{_1}+n_{_2}+n_{_3})}
{n_{_1}!n_{_2}!n_{_3}!\Gamma(2-{D\over2}+n_{_1})\Gamma(2-{D\over2}+n_{_2})\Gamma({D\over2}+n_{_3})}\;.
\label{3loop-S47-3}
\end{eqnarray}

\item $I_{_8}=[1,3,4,5,7]$, i.e.
the implement $J_{_8}=[1,8]\setminus I_{_8}=[2,6,8]$.
The choice implies the exponent numbers $\alpha_{_2}=\alpha_{_6}=\alpha_{_8}=0$, and
\begin{eqnarray}
\alpha_{_1}=\alpha_{_3}=\alpha_{_5}=1-\frac{D}{2},\;
\alpha_{_4}=D-3,\;\alpha_{_7}=\frac{3D}{2}-4\;.
\label{3loop-S48-1}
\end{eqnarray}
The corresponding hypergeometric series solution is written as
\begin{eqnarray}
&&\Phi_{_{[13\tilde{4}]}}^{(8)}(\alpha,z)=
y_{_2}^{\frac{3D}{2}-4}\sum\limits_{n_{_1}=0}^\infty
\sum\limits_{n_{_2}=0}^\infty\sum\limits_{n_{_3}=0}^\infty
{c_{_{[13\tilde{4}]}}^{(8)}(\alpha,{\bf n})}
\Big({1\over y_{_2}}\Big)^{n_{_1}}
\Big({y_{_1}\over y_{_2}}\Big)^{n_{_2}}
\Big({y_{_3}\over y_{_2}}\Big)^{n_{_3}}\;,
\label{3loop-S48-2}
\end{eqnarray}
with the coefficient is
\begin{eqnarray}
&&c_{_{[13\tilde{4}]}}^{(8)}(\alpha,{\bf n})=
\frac{\Gamma(4-{3D\over2}+n_{_1}+n_{_2}+n_{_3}) \Gamma(3-D+n_{_1}+n_{_2}+n_{_3})}
{n_{_1}!n_{_2}!n_{_3}!\Gamma(2-{D\over2}+n_{_1})\Gamma(2-{D\over2}+n_{_2})\Gamma(2-{D\over2}+n_{_3})}\;.
\label{3loop-S48-3}
\end{eqnarray}

\end{itemize}

\end{document}